\documentclass[preprint,12pt]{elsarticle}

\usepackage{moresize}
\usepackage{multirow}
\usepackage{listings}
\usepackage{hyperref}
\usepackage{caption}
\usepackage{lineno}
\usepackage{color}

\usepackage{amssymb}
\usepackage{amsmath}

\journal{Journal of Systems and Software}

\begin{document}

\begin{frontmatter}




\title{StoneDetector: Conventional and versatile \\ code clone detection for Java} 

\author[dhge]{Thomas S. Heinze} 

\affiliation[dhge]{organization={Cooperative University Gera-Eisenach},
            addressline={Weg der Freundschaft 4}, 
            city={Gera},
            postcode={07546}, 
            country={Germany}}

\author[fsu]{Andr\'e Sch\"afer} 
\author[fsu]{Wolfram Amme} 
\affiliation[fsu]{organization={Friedrich Schiller University Jena},
            addressline={Ernst-Abbe-Platz 2}, 
            city={Jena},
            postcode={07743}, 
            country={Germany}}

\begin{abstract}
Copy \& paste is a widespread practice when developing software and,
thus, duplicated and subsequently modified code occurs frequently in software
projects. Since such code clones, i.e., identical or similar fragments of
code, can bloat software projects and cause issues like bug or vulnerability
propagation, their identification is of importance. In this paper, we present
\textcolor{black}{StoneDetector} and its underlying method for finding code
clones in Java source and Bytecode. StoneDetector implements a conventional
clone detection approach based upon the textual comparison of paths derived
from the code's representation by dominator trees. In this way, the tool does
not only find exact and syntactically similar near-miss code clones, but also
code clones that are harder to detect due to their larger variety in the
syntax. We demonstrate StoneDetector's versatility as a conventional
clone detection \textcolor{black}{tool} and analyze its various available configuration
parameters, including the usage of different string metrics, hashing
algorithms, etc. In our exhaustive evaluation with other conventional clone
detectors on several state-of-the-art benchmarks, we can show StoneDetector's
performance and scalability in finding code clones in both, Java source
and Bytecode.
\end{abstract}


\begin{keyword}
code clone \sep clone detection \sep Java \sep static analysis
\end{keyword}

\end{frontmatter}

\section{Introduction} \label{sec:introduction}

\begin{figure}
  \ssmall
\begin{minipage}[t]{.9\textwidth}
\begin{lstlisting}[language=Java,showstringspaces=false]
int func(int[] array) {
    int sum = 0;
    for (int i = 0; i<array.length; i++)
        if (array[i]>0)
            sum *= array[i];
    return sum;
}

\end{lstlisting}
                  
(a) Original code fragment

\begin{lstlisting}[language=Java,showstringspaces=false]
int func( int[] array ) {
    int sum = 0;
    for ( int i = 0; i < array.length; i++ )
        if ( array[i] > 0 )
            sum *= array[i];
    return sum;
}
\end{lstlisting}
(b) Type 1 clone      
  
\begin{lstlisting}[language=Java,showstringspaces=false]
int func(int[] arr) {
    int s = 0;
    for (int i = 0; i<arr.length; i++)
        if (arr[i]>0)
            s *= arr[i];
    return s;
}

\end{lstlisting}
(c) Type 2 clone       

\begin{lstlisting}[language=Java,showstringspaces=false]
int func(int[] arr) {
    int s = 0;
    for (int i = 0; i<arr.length; i++)
        if (arr[i]>0)
            s *= (arr[i]*fu());
    return s;
}

\end{lstlisting}
(d) Type 3 clone
      
\begin{lstlisting}[language=Java,showstringspaces=false]
int func(int[] array) {
    int sum = 0;
    int i = 0;
    while (i<array.length) {
        sum = (array[i]>0)?sum*array[i]:sum;
        i = i+1;
    }
    return sum;
}

\end{lstlisting}
(e) Type 4 clone
\end{minipage}      
  \caption{Variations of code clones: Examples of Type 1, 2, 3, and 4 code clones}
	\label{fig:clones}
\end{figure}

\emph{Code clones}~\cite{DBLP:journals/scp/RoyCK09} are identical or similar
fragments of code that, e.g., arise due to copy-and-paste practices in
software development. Since such practices are quite common, code clones can
be frequently found, as empirical studies on code duplication have
shown in the past~\cite{DBLP:conf/wcre/RoyC08,DBLP:conf/iceccs/ZibranSAR11,DBLP:journals/pacmpl/LopesMMSYZSV17}.
While code cloning may initially speed up software development or allow for
reusing working code, it also introduces long-term challenges: Duplicated code
inflates code bases and increases the maintenance burden, as developers have
to manually track and update all copies for fixing bugs or implementing
code changes. The effects of code cloning are thus
controversial~\cite{DBLP:journals/ese/KapserG08,DBLP:conf/icse/JurgensDHW09}
and detecting code clones is a vital step toward maintaining code quality,
reducing technical debt, and improving program comprehension.

Code clones can come in different variations, ranging from exact replicas of
code, over code segments that have been modified in structure or syntax, to
fragments that have a completely different syntax but share the same 
computations (see also Fig.~\ref{fig:clones}).
In the literature, code clones are typically categorized into
four types based on their similarity~\cite{DBLP:journals/scp/RoyCK09}:
\emph{Type 1 clones}, which are exact duplicates of code fragments that only
differ in layout, comments, or formatting (cf. Fig.~\ref{fig:clones} (b)).
\emph{Type 2 clones} in addition to Type 1 clones include variations such as
identifier renaming or literal changes (cf. Fig.~\ref{fig:clones} (c)).
\emph{Type 3 clones} go another step further and allow for some inserted,
altered, or deleted statements (cf. Fig.~\ref{fig:clones} (d)). Finally,
\emph{Type 4 clones} are syntactically distinct but perform the same or
similar computations (cf. Fig.~\ref{fig:clones} (e)).
\emph{Code clone detection} has been actively
researched~\cite{DBLP:journals/scp/RoyCK09,DBLP:journals/jss/NasrabadiPRRE23,DBLP:journals/infsof/RattanBS13,DBLP:journals/access/AinBAAM19}
and the state-of-the-art approaches for clone detection perform quite
well at finding exact and syntactically very similar
code  clones, i.e., clones of Type 1-2 and part of
Type 3~\cite{DBLP:conf/iwpc/RoyC08a}. In contrast, detecting the
harder-to-find clone variations of Type 3 and Type 4, e.g.,
large-gap clones, which include many syntactical
modifications~\cite{DBLP:conf/icse/WangSWXR18}, or semantic clones,
which feature low syntactical similarity but share the same computations~\cite{DBLP:conf/sigsoft/SainiFLBL18},
is an open and challenging research problem. In this paper, we address this problem and
are interested in advancing the detection of code clones with larger
syntactical variance.

\begin{figure}
	\begin{minipage}[t]{\textwidth}
  \ssmall
\begin{lstlisting}[language=Java,showstringspaces=false,numbers=left]
private static void zipDir(File zipDir, ZipOutputStream zos, String path) {
  final String [] dirList = zipDir.list();
  final byte [] readBuffer = new byte[chunk];
  int bytesIn = 0;
  for (int i = 0; i < dirList.length; i ++) {
    final File file = new File(zipDir, dirList[i]);
    final String filePath =
            ((path != null && path.length() > 0) ? (path+File.separator) : "")
            + file.getName();
    if (file.isDirectory()) {
      zipDir(file, zos, filePath);
      continue;
    }
    final FileInputStream fis = new FileInputStream(file);
    try {
      final ZipEntry anEntry = new ZipEntry(filePath);
      zos.putNextEntry(anEntry);
      while ((bytesIn=fis.read(readBuffer)) != - 1) {
        zos.write(readBuffer, 0, bytesIn);
      }
      zos.flush();
    } finally {
      fis.close();
    }
  }
}
\end{lstlisting}
        \end{minipage}
        
	\begin{minipage}[t]{\textwidth}
  \ssmall
\begin{lstlisting}[language=Java,showstringspaces=false,numbers=left]
public static void zipDir(File zipDir, ZipOutputStream zos, String path) {
  String [] dirList = zipDir.list();
  byte [] readBuffer = new byte[2156];
  int bytesIn = 0;
  for (int i = 0; i < dirList.length; i ++) {
    File f = new File(zipDir, dirList[i]);
    if (f.isDirectory()) {
      zipDir(f, zos, path + f.getName() + "/");
    } else {
      FileInputStream fis = new FileInputStream(f);
      ZipEntry anEntry = new ZipEntry(path + f.getName());
      zos.putNextEntry(anEntry);
      while ((bytesIn=fis.read(readBuffer)) != - 1) {
        zos.write(readBuffer, 0, bytesIn);
      }
      fis.close();
    }
  }
}

\end{lstlisting}
  \end{minipage}
	  \caption{Example of two code fragments that are more difficult to identify as code clone}
	   \label{fig:zipdir}
\end{figure}

An example of such a harder-to-detect code clone is shown in
Fig.~\ref{fig:zipdir}, taken from the state-of-the-art clone detection
benchmark BigCloneBench~\cite{DBLP:conf/icsm/SvajlenkoR15,DBLP:conf/icsm/SvajlenkoR16}.
Conventional clone detectors have difficulties in identifying the code
clone due to the larger syntactical difference, e.g.,
NiCad~\cite{DBLP:conf/iwpc/RoyC08a,DBLP:conf/iwpc/CordyR11a} is only
able to determine the code clone accepting up to 58\% syntactical
difference, i.e., threshold value 0.58, which is considered to constitute
a Type 4 code clone according to the literature~\cite{DBLP:conf/icsm/SvajlenkoR15}.
In contrast, our proposed method and clone detection tool, StoneDetector,
finds the code clone with a similarity threshold of just 0.26.
A closer look at the example reveals that the larger deviation 
between both code fragments is to a large extent due to different 
syntactical constructs implementing similar control flow.
For instance, the {\bfseries\lstinline[columns=fixed]{continue}} in line 12 of
the upper code fragment in Fig.~\ref{fig:zipdir} is replaced by the
{\bfseries\lstinline[columns=fixed]{else}} branch in the lower code fragment
and the {\bfseries\lstinline[columns=fixed]{try-finally}} in lines 15-24 of the
upper fragment is replaced by an instruction sequence with the
{\bfseries\lstinline[columns=fixed]{finally}} block as the sequence's last
instruction in the lower fragment of Fig.~\ref{fig:zipdir}. In order
to find such harder-to-detect code clones, we present in this paper a
method for clone detection working on dominator trees. This method
exploits the semantics of the code fragments' control flow for a
more permissive textual comparison by encoding and comparing
paths of the code fragments' dominator trees. Dominator trees are a
traditional code representation, used in compilers or program analysis,
that provides us with a more informed code abstraction when compared to
the raw code, token sequences, or syntax trees as used in other clone
detectors. As a consequence, the method is able to provide better
performance in the detection of code clones with larger syntactical
variance, i.e., Type 3 and Type 4 code clones.

The presented clone detection method is implemented in \textcolor{black}{StoneDetector},
which is publicly available\footnote{\url{https://stonedetector.fmi.uni-jena.de/}}
and allows for finding code clones in Java source and Bytecode. Using
\textcolor{black}{StoneDetector} enabled us to perform an in-depth evaluation of the
method's performance on various state-of-the-art benchmarks,
including BigCloneBench~\cite{DBLP:conf/icsm/SvajlenkoR15,DBLP:conf/icsm/SvajlenkoR16},
Google Code Jam, Project CodeNet~\cite{DBLP:conf/nips/Puri0JZDZD0CDTB21},
and GPTCloneBench~\cite{DBLP:conf/icsm/AlamRARRS23}. As a result, we
found that our clone detection method and its implementation in 
\textcolor{black}{StoneDetector} allow for finding exact and near-miss code
clones in Java source code at competitive performance with other conventional
code clone detectors, including NiCad~\cite{DBLP:conf/iwpc/CordyR11a},
iClones~\cite{DBLP:conf/csmr/GodeK09},
SourcererCC~\cite{DBLP:conf/icse/SajnaniSSRL16},
Deckard~\cite{DBLP:conf/icse/JiangMSG07},
CCAligner~\cite{DBLP:conf/icse/WangSWXR18},
NIL~\cite{DBLP:conf/sigsoft/NakagawaHK21},
\textcolor{black}{CloneWorks~\cite{DBLP:conf/icse/SvajlenkoR17a}}
and Oreo~\cite{DBLP:conf/sigsoft/SainiFLBL18}. Additionally, StoneDetector
in particular excels at finding source code clones with larger syntactical
variance at high precision. StoneDetector also scales to large codebases
comprising multiple 100 million lines of code and features runtimes
comparable to the state of the art. Furthermore, we have experimented
with different configuration parameters for clone detection, e.g.,
threshold value, code clone size, applied string metrics, and hashing
algorithms, etc., and thereby show the versatility of \textcolor{black}{StoneDetector}
besides identifying its optimal configuration. Eventually,
we analyzed the method's potential in also finding code clones in
compilation artifacts like Java Bytecode, in cases where the source code
is unavailable in the first place. In summary, the contributions of this
paper are:
\begin{itemize}
  \item We demonstrate the feasibility of finding code clones in Java
        source code fragments by encoding and textually comparing paths
        of the code fragments' dominator trees. 
  \item We provide the publicly available \textcolor{black}{tool StoneDetector} with
        an implementation of our code clone detection method as
        a versatile conventional code clone detector and evaluate
        its performance and scalability.
  \item In its evaluation, we can validate that StoneDetector finds
        exact and near-miss Java source code clones at competitive
        recall and precision and is able to find even more of the
        harder-to-find clones with larger syntactical variance,
        when compared to other conventional code clone detectors
        on state-of-the-art benchmarks.
  \item We provide an analysis of the usage of different string
        metrics for clone detection within StoneDetector, considering
        string metrics like Levenshtein, Needleman-Wunsch, Hamming
        distance, and longest common subsequence (LCS)-based metric.
  \item We describe a variant of StoneDetector, which in particular
        performs well on large-gap code clones by using a combination
        of a modified LCS metric with locality sensitive hashing (LSH).
  \item We study the role of various additional configuration parameters,
        e.g., threshold value, minimal size of code clones, etc.,
        on clone detection with StoneDetector to empirically derive
        its optimal configuration. 
  \item We provide a variant of StoneDetector that works on Java
        Bytecode and experiment with its stack- and register-based
        intermediate representations.
\end{itemize}

This work is based upon the authors' prior conference paper on code clone
detection presented at the
\emph{2021 IEEE International Conference on Software Maintenance and
Evolution}~\cite{DBLP:conf/icsm/AmmeHS21} and extends it in various ways.
First, an analysis of the usage of different string metrics within StoneDetector
has been added to the paper, including distance metrics like Levenshtein,
Needleman-Wunsch, Hamming distance, and longest common subsequence (LCS)-based
metric. Among others, the analysis revealed the versatility of StoneDetector
as a clone detection framework: For example, using LCS provided the best
tradeoff in terms of recall, precision, and scalability. In contrast,
using the Hamming distance yielded a very fast clone detector with impaired
performance in finding harder-to-detect Type 3/4 code clones. Second, we
now derive the optimal configuration of StoneDetector for clone detection
by means of an in-depth empirical analysis of its several configuration
parameters (threshold value for path comparison, etc.). Furthermore,
expanding the approach for detecting large-gap and subclones, as presented
in our previous conference paper~\cite{DBLP:conf/icsm/AmmeHS21}, we added
locality-sensitive hashing (LSH) and a modified LCS metric to StoneDetector,
which resulted in performance gains for these types of code clones.
We now also include a more exhaustive analysis of StoneDetector's performance
in finding harder-to-detect Type 3/4 code clones by including three more
state-of-the-art benchmarks for code clone detection besides BigCloneBench,
i.e., Google Code Jam, Project CodeNet, and GPTCloneBench, and \textcolor{black}{three} more
clone detectors, i.e., iClones, \textcolor{black}{CloneWorks~\cite{DBLP:conf/icse/SvajlenkoR17a}},
and NIL, in the evaluation. Moreover, we
also extended our previous approach in~\citet{DBLP:conf/icsm/AmmeHS21}
for clone detection on stack-based and register-based Java Bytecode
representations, such that \textcolor{black}{StoneDetector} can be used for
both finding code clones in Java source code and Bytecode.

The rest of the paper is structured as follows:
\textcolor{black}{Sect.~\ref{sec:implementation}
introduces \textcolor{black}{StoneDetector}'s overall approach and architecture. The
underlying clone detection method based on encodings of paths
in the dominator tree and string metrics for path comparison are described
in detail in Sect.~\ref{sec:approach}.} In 
Sect.~\ref{sec:evaluation}, we present the in-depth evaluation of our
approach on various benchmarks for Java code clone detection, including
the comparison with state-of-the-art clone detectors
and the thorough analysis of \textcolor{black}{StoneDetector}'s various
configuration parameters. Related work on code clone detection is
discussed in Sect.~\ref{sec:relatedwork} and Sect.~\ref{sec:conclusion}
finally concludes the paper \textcolor{black}{and mentions future work.}

\section{\textcolor{black}{Overall Approach and Architecture}} \label{sec:implementation}

\textcolor{black}{Our approach to code clone detection has been implemented in
\emph{\textcolor{black}{StoneDetector}}~\cite{DBLP:conf/icsm/AmmeHS21,DBLP:conf/iwsc/SchaferAH23}
thus providing a full-scale stand-alone clone detector for Java.} As input,
StoneDetector expects a number of Java source or Bytecode files for which the
\textcolor{black}{tool} generates a list of all code clone pairs, which are identified and mapped
back to the original code by file location and line numbers. The resulting
information can then be used, for example, to search for similar programs in code
repositories~\cite{DBLP:conf/iwpc/Keivanloo12}, to analyze the evolution of
code~\cite{DBLP:journals/tse/NguyenNPAN12}, or to track bugs and vulnerabilities
propagated via duplicated code~\cite{DBLP:journals/tse/LiLMZ06}.

\begin{figure}[t]
  \centering
  \includegraphics[width=\textwidth]{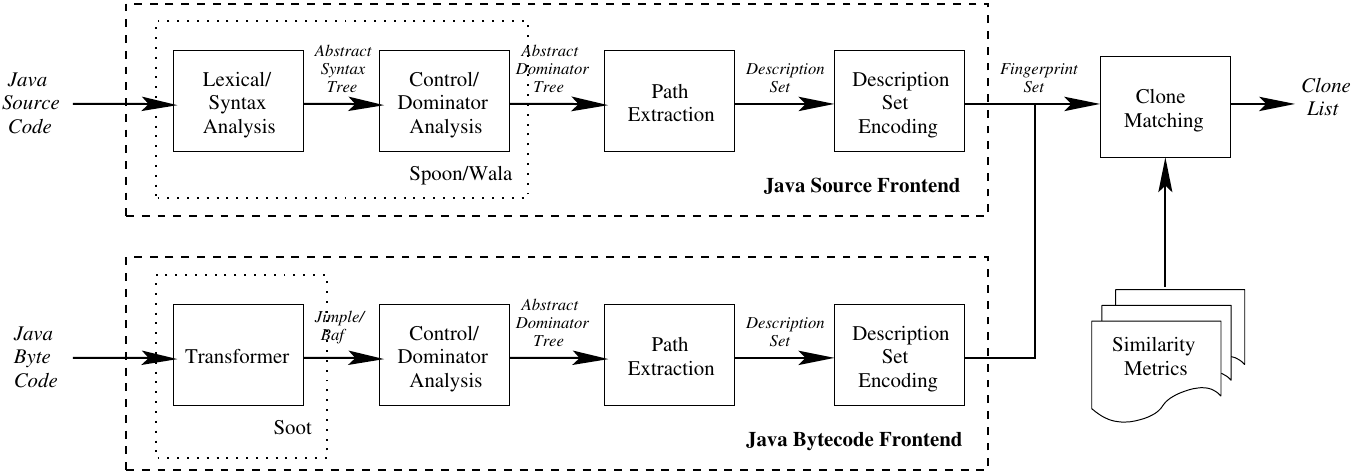}
  \caption{\textcolor{black}{Overview of the tool StoneDetector}}
  \label{fig:system}
\end{figure}

An overview of \textcolor{black}{StoneDetector} can be seen in Fig.~\ref{fig:system},
which shows its inner structure and how its components interact with
each other. Considering, to begin with, Java source code, the \textcolor{black}{tool} starts
with performing lexical and syntax analysis on the given Java source files.
This step yields the abstract syntax trees, which are subsequently subject to
control flow analysis for generating (abstract) dominator trees. Afterward,
paths are extracted from the generated dominator trees and linearized into
description sets. The individual nodes of
the description sets can be further encoded using a hashing algorithm, which
yields fingerprint sets. Either raw description sets or fingerprint sets are
eventually transferred to the clone matcher, which checks for similarities
using one out of the several available string metrics. Based upon this similarity
analysis, code clones are identified and returned for further processing.
We will next provide more information on the individual components of the
described processing pipeline for Java source code files. \textcolor{black}{For
a more elaborative introduction of the clone detection method based on
dominator tree and string metrics we refer the reader to Sect.~\ref{sec:approach}.}

\paragraph*{Lexical/Syntax Analysis}
The lexical and syntax analysis in the Java source code frontend of
\textcolor{black}{StoneDetector} is utilizing the
\emph{Spoon analysis framework}~\cite{pawlak:hal-01169705}\footnote{\url{https://spoon.gforge.inria.fr/}}. 
The Spoon framework includes libraries for parsing Java source code
files up to Java version 20 into abstract syntax trees (AST) and provides the
AST metamodel and functionality for manipulating them.

\paragraph*{\textcolor{black}{Control/Dominator Analysis}}
The Spoon framework also supports generating control flow graphs of
individual Java functions based on its AST model. We can therefore resort to
Spoon's control flow builder for our control flow analysis. However,
the framework currently includes neither a meta-model for
dominator trees nor any dominator analysis. For the construction of
dominator trees, the respective builder from the
\emph{WALA framework}~\cite{wala}\footnote{\url{https://github.com/wala/WALA}}
has been integrated into \textcolor{black}{StoneDetector}. As before, dominator
trees are built at the function level. In addition, due to the coarse
abstraction for Java's exception handling in control flow graphs generated
by Spoon, we have implemented a standard interprocedural exception
analysis for Java as described by~\citet{Chang01interproceduralexception}.
Using the analysis allows for a more fine-grained control flow mapping of
Java's exception handlin and experiments
have shown that enabling exception analysis implies better results for
clone detection. 

\paragraph*{Path Extraction and Description Set Encoding} 
Dominator trees are transformed
into description sets in order to subsequently perform textual clone matching.
To this end, a (abstract) dominator tree is traversed such that the set of its
paths starting at the root node and ending at a leaf nodes, respectively, is
generated. Note that we in general perform identifier and
literal abstraction in the resulting description sets, much like as in other clone
detectors, e.g., NiCad~\cite{DBLP:conf/iwpc/RoyC08a}. Though function names at
call sites may optionally be conserved, which provided better experimental results
in some cases. We therefore offer a switch for enabling/disabling identifier
abstraction for function names at call sites. Likewise, \textcolor{black}{StoneDetector}
treats encoding of description sets into fingerprint sets via a hashing
function as optional. Raw description sets can thus be directly committed to the
clone matcher or beforehand transformed such that the contained instructions
are encoded using one of the several implemented hashing algorithms, i.e.,
MD5, 4-byte prime number hashing, or \textcolor{black}{Locality-Sensitive Hashing (LSH)}.

\paragraph*{Clone Matching} The clone matching component receives a set of
description or fingerprint sets as input and performs pairwise similarity
checks. The similarity of
description sets, i.e., code fragments, is traced back to the accumulated
similarity of its paths in our approach. Path similarity is computed using
one out of several string metrics available in \textcolor{black}{StoneDetector}.
Among the currently available metrics are the Hamming
distance~\cite{DBLP:books/daglib/0068879}, the
Levenshtein distance~\cite{DBLP:journals/iandc/Levenshtein75}, the
Needleman-Wunsch-based metric~\cite{Needleman1970}, as well as the
LCS-based metric~\cite{DBLP:conf/spire/BergrothHR00}.
Further supported metrics include the N-Gram, Jaccard, Cosine, Jaco-Winkler,
and Damerau-Levenshtein metrics. 

The thus defined processing pipeline in \textcolor{black}{StoneDetector} is parallelized.
On the one hand, generation of description sets can be done in parallel with the
help of streams. If all description sets have been eventually generated, clone
matching begins. This process, on the other hand, can also be done in parallel,
again using streams. The configuration of a specific run of \textcolor{black}{StoneDetector}
allows for the specification of the number of threads to be used in
both computations. 
In order to further increase throughput, especially when processing larger
numbers of code fragments, a copy strategy can optionally be used for description set
handling. NiCad uses a similar technique for improving its runtime
behavior~\cite{DBLP:conf/iwpc/RoyC08a,DBLP:conf/iwpc/CordyR11a}. In
StoneDetector, however, the technique has been adopted in such a way that
the number of identified clones remains unchanged whether or not the
technique is enabled. The basic idea of the copy strategy is to
continue with further comparisons only for one of the two code fragments $f$ and $g$
whenever a clone pair $f$ and $g$ is found: Suppose comparisons are continued
for fragment $f$ and terminated for fragment $g$. After all comparisons are
finished, code clones found for $f$ are added to the clones already found for
$g$, up to termination. Note that in
our variant of the copy strategy, two code fragments are only considered as
a clone pair if they are identical to each other, i.e., if they have a distance
value of zero. Using our copy strategy, runtime behavior does not improve as
much as in NiCad, however, accuracy of the resulting code clones is guaranteed.

Apparently, as can be seen by the above discussion, \textcolor{black}{StoneDetector}
provides a rich set of configuration parameters that can be used for
optimizing clone detection and demonstrates the \textcolor{black}{tool's} versatility. Among
the most critical parameters are the threshold value $\tau$ defining the
required similarity of code clones, the minimum size $l$ of code fragments
acceptable as code clones, the string metric $m$ used for path
matching, and the hash function \textit{h} used for description set encoding
(cf. also \textcolor{black}{Table~\ref{fig:options}}). \textcolor{black}{The therein
given default values of configuration parameters are chosen for the
tool's best tradeoff of recall and precision with acceptable runtime.}
For an in-depth description of the \textcolor{black}{StoneDetector}'s configuration
parameters, we refer the reader \textcolor{black}{to the evaluation in
Sect.~\ref{sec:evaluation} and} to the supplementary information available
online\footnote{\url{https://stonedetector.fmi.uni-jena.de/}}.

\begin{table}
 \footnotesize
\centering
\begin{tabular}[h]{l|l|l}
\emph{Configuration parameter}&\emph{Value range}&\emph{Default value}\\
\hline
Similarity threshold $\tau$ &  $0\le \tau \le 1.0$ & $0.3$ \\ 
\hline
Minimum clone size in code lines $l$ & $0\le l $&15\\
\hline
String metric $\delta$ & LCS & \textcolor{black}{LCS$_{\mathrm{modified}}$}\\
& LCS$_{\mathrm{modified}}$ &\\
&Levenshtein&\\
&Needleman-Wunsch&\\
&Hamming&\\
&N-Gram&\\
&Jaccard&\\
&Cosinus&\\
&Jaco-Winkler&\\
&Damerau-Levenshtein&\\
\hline
Hashing algorithm $h$ & none & \textcolor{black}{LSH}\\
&4-byte hashing& \\
&MD5& \\
&LSH& \\
\end{tabular}
\captionof{table}{\textcolor{black}{Main configuration parameters in \textcolor{black}{StoneDetector}}}
  \label{fig:options}
\end{table} 

\subsection{Extensibility of \textcolor{black}{StoneDetector}}

The use of fingerprints to represent description sets makes clone detection in
\textcolor{black}{StoneDetector} fundamentally independent of the selected programming language
and possible program representation. In fact, as is also depicted in Fig.~\ref{fig:system},
StoneDetector can be divided into a frontend and backend, with the former generating
description sets of given code fragments and the latter performing the actual clone matching.
The strict separation of frontend and backend allows us to facilitate a readily extensible
\textcolor{black}{tool}, which can be easily extended for processing other programming languages as well
as other internal program representations, such as control flow graphs, abstract syntax
trees, or even just token sequences of linear code.
Such an extension is only required to map the new programming language or program representations
to fingerprints by using a certain hash function. Clone matching can subsequently be performed
using the existing backend. The specific hash function used for
construction is irrelevant to the functionality of the backend. In this way, fingerprints form
a general and elegant interface for extending our system to other programming
languages and program representations.

To demonstrate the \textcolor{black}{StoneDetector}'s versatility and extensibility, we have added the handling of
Java Bytecode as another frontend of \textcolor{black}{StoneDetector}~\cite{DBLP:conf/iwsc/SchaferHA23}
(cf. Fig.~\ref{fig:system}).
The new frontend gets a set of class files as input and transforms them to fingerprint sets,
which are committed to the existing clone matcher of StoneDetector.
Specifically, processing class files is implemented via the
\emph{Soot framework}~\cite{DBLP:conf/cascon/Vallee-RaiCGHLS99}, which first transforms the contained
Bytecode into the Bytecode-like \emph{Baf} or into the register-based \emph{Jimple} representation.
Afterward, dominator trees are generated based upon the Baf or Jimple representation and
paths contained in the dominator trees are extracted and encoded into 
fingerprint sets, as required for clone matching.  

\section{\textcolor{black}{Clone Detection using Dominator Trees}} \label{sec:approach}

The clone detection method described in this paper performs textual
comparisons of code fragments' syntax similar to other text-based or
lexical approaches for code clone detection, though the method
additionally employs comparisons of the fragments' control flow
at a higher level of abstraction. Moreover, we do not use
abstract syntax trees or control flow graphs for this purpose, as in
other tree- or graph-based approaches, but the more specialized code
representation of dominator trees. Dominator trees~\cite{DBLP:books/aw/AhoSU86}
are usually utilized in compilers to represent the execution order of
instructions. They feature the advantage of, on the one hand, modeling
control flow on a higher level when compared to plain syntax or
abstract syntax trees. On the other hand, dominator trees support
straightforward linearization due to the fact that each node is
reachable via one unambiguous path and due to the absence of cycles, in
contrast to control flow or program dependence graphs.

\subsection{Control flow and Dominator Trees}

\begin{figure}
  \flushleft
  \ssmall
\footnotesize
\begin{minipage}{0.45\textwidth}
\centering
\includegraphics[scale=.3]{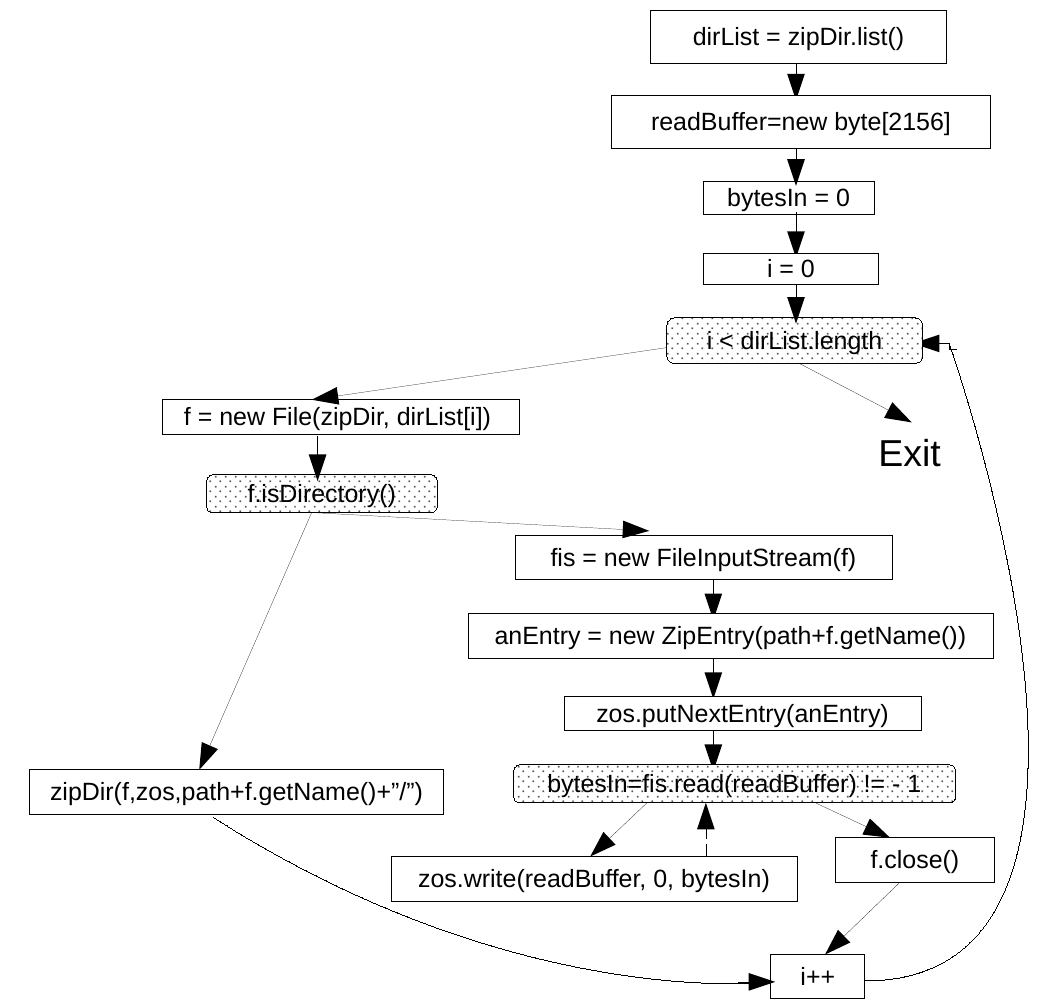}\\
\centering  (a) Control Flow Graph
\end{minipage}
\begin{minipage}{0.45\textwidth}
\centering
\includegraphics[scale=.28]{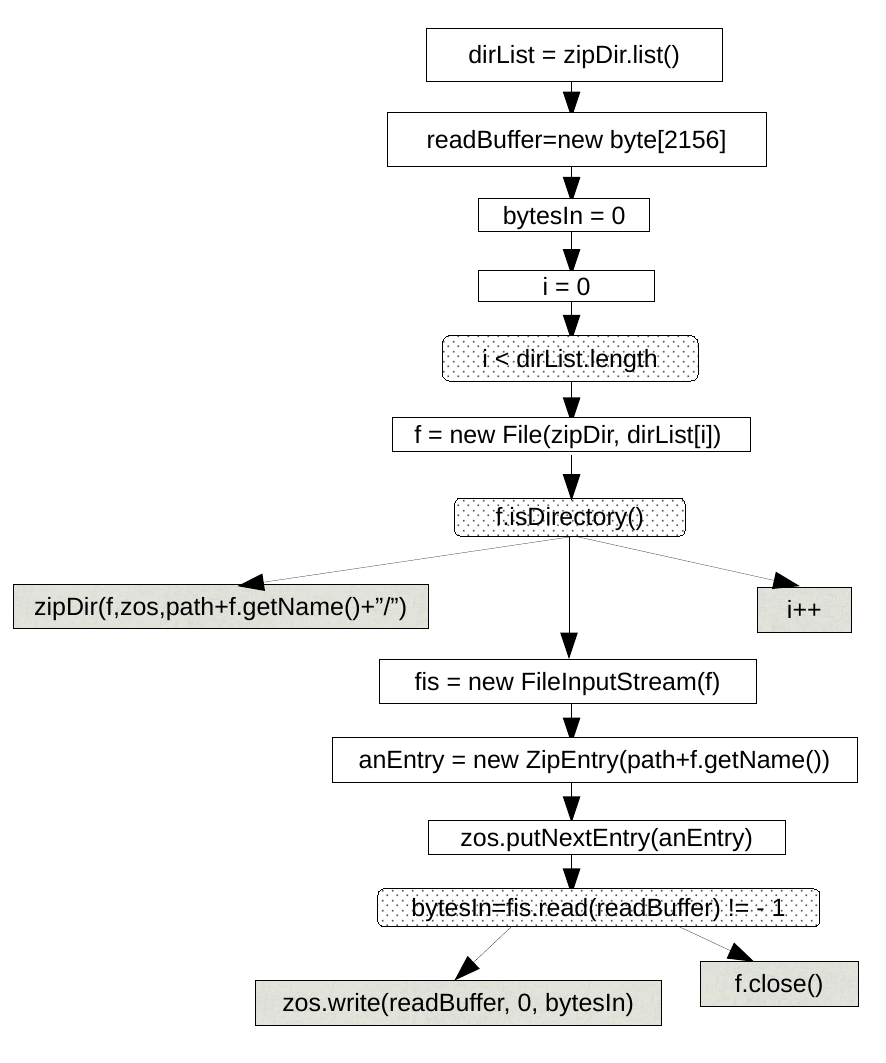}\\
\centering  (b) Dominator Tree
\end{minipage}\\
\begin{minipage}{0.65\textwidth}
  \centering
  \includegraphics[scale=.27]{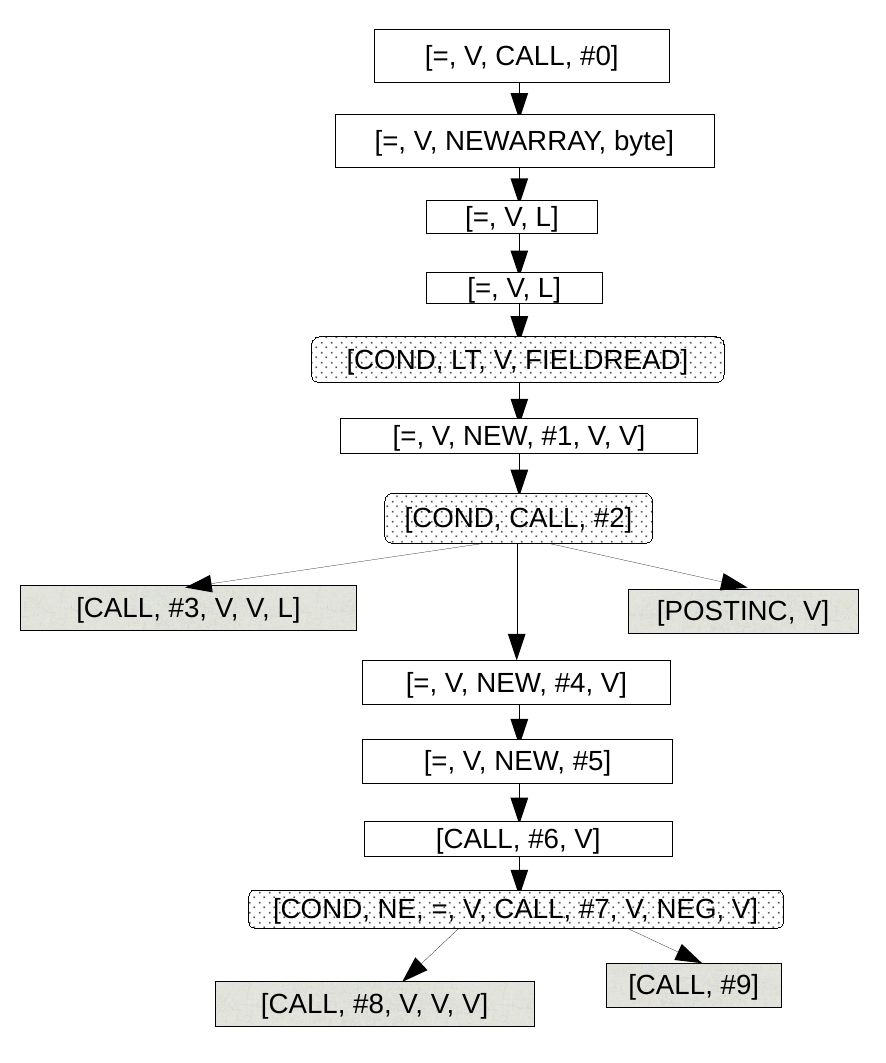}
\vspace{0.2cm}
\centering (c) Abstract Dominator Tree
\end{minipage} 
\begin{minipage}[c]{0.3\textwidth}
    \begin{center}\includegraphics[scale=0.4]{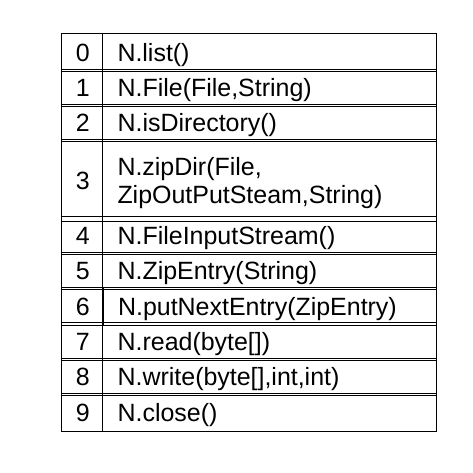}\end{center}
\end{minipage}\\
\begin{minipage}{\textwidth}
  \ssmall
\begin{verbatim}
Dset= [[=, V, CALL, #0]->[=, V, NEWARRAY, byte]->[=, V, L]->[=, V, L]->[COND, LT, V, FIELDREAD]->
       [=, V, NEW, #1, V, V]->[COND, CALL, #2]->[CALL, #3, V, V, L],
       [=, V, CALL, #0]->[=, V, NEWARRAY, byte]->[=, V, L]->[=, V, L]->[COND, LT, V, FIELDREAD]->
       [=, V, NEW, #1, V, V]->[COND, CALL, #2]->[=, V, NEW,  #4, V]->[=, V, NEW, #5]->
       [CALL, #6, V]->[COND, NE, =, V, CALL, #7, V, NEG, V]->[CALL, #8, V, V, V],
       [=, V, CALL, #0]->[=, V, NEWARRAY, byte]->[=, V, L]->[=, V, L]->[COND, LT, V, FIELDREAD]->
       [=, V, NEW, #1, V, V]->[COND, CALL, #2]->[=, V, NEW,  #4, V]->[=, V, NEW, #5]->
       [CALL, #6, V]->[COND, NE, =, V, CALL, #7, V, NEG, V]->[CALL, #9],
       [=, V, CALL, #0]->[=, V, NEWARRAY, byte]->[=, V, L]->[=, V, L]->[COND, LT, V, FIELDREAD]->
       [=, V, NEW, #1, V, V]->[COND, CALL, #2]->[POSTINC, V]]
\end{verbatim}
\vspace{-0.7cm}
\begin{verbatim}
Fingerprints=
[35d37480aba8b098914b104b822d27af->d352f27168825f06297c0968af4006cc->c0812f2a5a758d7f6572a0f8fc057ce3->
 c0812f2a5a758d7f6572a0f8fc057ce3->20ae2bed041f26248c27148484197862->71bd8588d1df5ed91c8d27fb91a55d46->
 a45688ae3fc84165d19f2efe859e43db->c088b1671e63a009662f8e31c773a0a0,
 35d37480aba8b098914b104b822d27af->d352f27168825f06297c0968af4006cc->c0812f2a5a758d7f6572a0f8fc057ce3->
 c0812f2a5a758d7f6572a0f8fc057ce3->20ae2bed041f26248c27148484197862->71bd8588d1df5ed91c8d27fb91a55d46->
 a45688ae3fc84165d19f2efe859e43db->c21875dc52243cb4f19c52d3e02be40a->19b4a2a27bfa0ccbeeee5564571ec011->
 cb4b446eaed074420c4e9c663ffd3741->c15ffb85b338cca72f6729adcc304469,
 35d37480aba8b098914b104b822d27af->d352f27168825f06297c0968af4006cc->c0812f2a5a758d7f6572a0f8fc057ce3->
 c0812f2a5a758d7f6572a0f8fc057ce3->20ae2bed041f26248c27148484197862->71bd8588d1df5ed91c8d27fb91a55d46->
 a45688ae3fc84165d19f2efe859e43db->055bfda33fe11d9fb52471e283c27d9c->19b4a2a27bfa0ccbeeee5564571ec011->
 cb4b446eaed074420c4e9c663ffd3741->f13c29192ec234b0a51eb4f32bc85efe,
 35d37480aba8b098914b104b822d27af->d352f27168825f06297c0968af4006cc->c0812f2a5a758d7f6572a0f8fc057ce3->
 c0812f2a5a758d7f6572a0f8fc057ce3->20ae2bed041f26248c27148484197862->71bd8588d1df5ed91c8d27fb91a55d46->
 a45688ae3fc84165d19f2efe859e43db->fc6ebb514a9a13b92cff0ae97894c23e]
\end{verbatim}
\end{minipage}\\
 \vspace{0.2cm}
 \centering (d) Description Set and Fingerprints
 \vspace{-0.2cm}
	\caption{Control flow graph, dominator tree, description set, and fingerprints of example}
        \label{fig2}
\end{figure}

A \emph{control flow graph} $G$ is a directed graph defined as
$G=(N,E,s)$ with $(N,E)$ denoting the graph and
$s\in N$ the start node~\cite{DBLP:books/aw/AhoSU86}. We encode
the instructions of a code fragment using the graph's nodes.
Note that we here use a single node for each instruction
instead of using basic blocks. An edge $(k,l)\in E$ in a control
flow graph connects two nodes $k$ and $l$ iff the instruction
denoted by $l$ can be immediately executed after the
instruction denoted by $k$. Each node $n\in N$ is then
reachable via a path from the start node $s$, i.e., a sequence
of consecutive edges $e\in E$ that begin at $s$
and end at $n$. 
An example of a control flow graph representing the lower code
fragment shown in Fig.~\ref{fig:zipdir}, i.e.,
function \lstinline[columns=fixed]{zipDir},
can be found in Fig.~\ref{fig2} (a).

A \emph{dominator tree}~\cite{DBLP:books/aw/AhoSU86} is a tree
and models a certain control flow property of code fragments,
specifically, which instruction precedes the other under all
possible executions of the respective code fragment. Thus,
let the triple $G=(N, E, s)$ denote a certain control flow
graph. A node $l\in N$ \emph{dominates} a node $k\in N$
iff $l$ occurs on all paths from the start node $s$ to node
$k$. Accordingly, every node $n\in N$ dominates itself by
definition. A node $l\in N$ furthermore
\emph{strictly dominates} another node $k\in N$ iff 
$l$ dominates $k$ and $l\neq k$. Node $l$ is also
\emph{immediately dominating} $k$ iff $l$ strictly dominates
$k$ and there exists no node $m\in N$, such that $m$
strictly dominates $k$ and $l$ strictly dominates $m$.
The dominator tree for the control flow graph $G$ is then
a directed graph $D=(N, E^*)$ with
$E^* =\{(l, k)\,|\,l \mbox{ immediately dominates }k\}$.
An example of a dominator tree is visualized in Fig.~\ref{fig2} (b).
As can be seen, the shown dominator tree corresponds to the control
flow graph in Fig.~\ref{fig2}(a). Note that in the dominator tree,
each node is reachable from the start node, i.e., the tree's
root, via a single unambiguous path. This is in contrast to the
given control flow graph. Also note that there are no backward edges 
encoding loop edges, and therefore cycles, in the dominator tree,
since the loop body cannot dominate the loop condition
according to the above definition.  

We use an abstract, text-oriented encoding of instructions for our
clone detection method, yielding the \emph{abstract dominator tree}
which is depicted in Fig.~\ref{fig2} (c). Therein, instructions and their
respective operands are denoted in preorder notation. Furthermore,
the distinct variable and field names, as well as literals, are not
distinguished and mapped to a single joint symbol, respectively.
Thus, all variable names are subsumed by symbol $V$, literals by
symbol $L$, and so on, much like in the clone detector
NiCad~\cite{DBLP:conf/iwpc/RoyC08a}. For example, consider
instruction \lstinline[columns=fixed]{i = 0} which is encoded by
\texttt{[=,V,L]} in the abstract dominator tree (cf. Fig.~\ref{fig2}).
Function calls are optionally represented using a constant pool.
A constant pool access is denoted by the character \texttt{\#} and the
respective function's entry in the constant pool, as, e.g., in
in \texttt{[CALL,\#6,V]} for
instruction \lstinline[columns=fixed]{zos.putNextEntry(anEntry)}
(again, cf. Fig.~\ref{fig2}). By using abstract dominator trees,
we are able to more easily match code fragments that do not
completely align with each other syntactically, in particular
in the case of Type 2 and part of Type 3/4 code clone deviations.

\subsection{Finding Similar Code Fragments} \label{sec:similarity}

Finding matching code fragments by using graph matching algorithms
on their respective dominator trees implies expensive computational
costs and is rather inefficient~\cite{DBLP:conf/ccs/XuLFYSS17}.
Instead of using a tree- or graph-based comparison of
dominator trees, we have chosen a text-based approach. Accordingly,
we linearize the dominator trees into sets of node sequences and,
to this end, derive the set of paths starting at a dominator tree's
root node, which are subsequently encoded into sequences
and used for the comparison.

Our approach is similar to the method proposed by~\citet{DBLP:journals/pacmpl/AlonZLY19},
where they use a linearization
of abstract syntax trees based on paths that start and end at distinct
leaf nodes of an abstract syntax tree. Though, in contrast, in our
method, given a dominator tree $D=(N,E^*)$, we consider all paths
starting at the tree's root node $s\in N$ and ending at a
respective leaf node $l\in \{n\in N\,|\,\nexists (n,m)\in E^*\}$. The
resulting set of these paths can then be used to unambiguously represent
the dominator tree $D$. A single path is encoded by simply concatenating
the (abstract) instructions denoted by the path's nodes into a sequence
of instructions. For example, considering the abstract dominator tree in
Fig.~\ref{fig2} (c), the path from the start node to the rightmost leaf
node is encoded by the sequence:
\smallskip \\
\indent \texttt{\footnotesize [=,V,CALL,\#0]$\rightarrow$[=,V,NEWARRAY,byte]$\rightarrow$[=,V,L]$\rightarrow$[=,V,L]
\\ \indent $\rightarrow$[COND,LT,V,FIELDREAD]$\rightarrow$[=,V,NEW,\#1,V,V]$\rightarrow$[COND,CALL,\#2]$\rightarrow$[POSTINC,V]}
\smallskip

Let $D$ be a dominator tree, $\mathit{paths}_{D}$ the set of its paths,
and $\mathit{leaves}_{D}$ the set of its leaf nodes in the following.
The set of paths unambiguously representing $D$ is given by:
$D_{\mathit{set}} = \{s\rightarrow\cdots\rightarrow n_k\, |\, s \cdots n_k \in \mathit{paths}_{D}, n_k \in \mathit{leaves}_{D}\}$.
We call the set $D_{\mathit{set}}$ the \emph{description set} of
dominator tree $D$. In the case of the abstract dominator tree depicted in
Fig.~\ref{fig2} (c), we have four paths starting at the start node and
ending in the leftmost, the left inner, the right inner, and the rightmost
leaf node, respectively. Accordingly, the description set shown in
Fig.~\ref{fig2} (d) contains the four instruction sequences representing
the paths.

Given two dominator trees $D$ and $D'$, which include two code fragments,
$f$ and $g$, respectively, we call $f$ and $g$ \emph{strictly similar} iff
both description sets $D_{\mathit{set}}$ and $D'_{\mathit{set}}$ have the
same number of elements and there exists a
similar path $p' \in D'_{\mathit{set}}$ for each path $p \in D_{\mathit{set}}$.
In cases where only the latter condition is satisfied, meaning that
$D_{\mathit{set}}$ and $D'_{\mathit{set}}$ do not share the same number of
elements, $f$ and $g$ are called \textit{partially similar}.

Note that we have not defined the similarity of paths, though. The similarity
of two paths included in the dominator trees $D$ and $D'$, respectively, 
can be formally assessed by a distance function
$\delta: paths_{\mathit{D}} \times paths_{\mathit{D'}} \rightarrow \mathbb{N}$ 
determining the distance of two paths. Basically, the distance function
$\delta$ can be freely chosen according to its requirements. The function
only has to satisfy the metric conditions, i.e., returning zero in case 
of identical paths, a value greater than zero for distinct paths, being symmetrical
in its arguments, and fulfilling the triangle inequality. The distance function
is later used as a metric for path similarity of dominators tree
and subsequently also for assessing the similarity of code fragments and
finding code clones.

\subsection{Metrics for Path Comparison} \label{sec:metrics}

For a na\"ive comparison of two paths $p$ and $q$ of dominator trees,
we could simply analyze their respective nodes, i.e., instructions,
for textual equality in the order of their occurrence. Such a comparison,
though, assumes that only those paths are similar that share the same
nodes; in other words, paths in which contained instructions are textually
identical. Thus, a more permissive approach is required, in particular
in the case of Type 3 and Type 4 code clones, where the paths of the dominator
trees may deviate from one another in some of their respective nodes. Even more,
a path may be embedded in the other matching path in case of large-gap or
subclones~\cite{DBLP:conf/icsm/AmmeHS21} 

A more permissive approach can be implemented
using string metrics, which are used in approximative string matching
and are therefore excellent candidates for our above-defined distance
function $\delta$. Since string metrics are based on text strings,
we depict the nodes contained in a path as individual characters, and two
nodes are then considered the same character if they represent the same
instruction. As a result, we can represent a single path of an (abstract)
dominator tree as a text string by concatenating the respective characters
of the path's nodes in the order of their occurrence. String metrics
are then readily applicable for the path comparison.

The \emph{Hamming distance}~\cite{DBLP:books/daglib/0068879} is a basic
string metric, which is defined on text strings with fixed length.
This distance measures the minimum number of substitutions to change
one string into another and thus yields the number of deviating
characters in two given strings. Therefore, as a limitation, the text
strings are only compared up to the length of the respective shorter
string, and the number of remaining characters in the longer string is
merely added to the so far calculated distance measure. Note that the
conventional Hamming distance therefore tends to penalize paths
with different lengths. \textcolor{black}{Note that the Hamming distance
can be computed quite efficiently, which makes it a good candidate
for fast and scalable near-miss code clone detection, as we have also observed
in our experiments (cf. Sect.~\ref{sec:evaluation}).}

\begin{figure}
  {
    \scriptsize
    \vspace{0.2cm}\textbf{Example:}
    \begin{itemize}
    \item[] Code fragment 1: for (int j=0; j$<$m; j++) \{ acc=acc-j; proc(acc); \} \\
            \hspace*{1.5in}\includegraphics[width=0.02\linewidth]{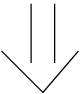}
     \vspace*{-.4cm}
   \item[] $f=\texttt{[=,V,L]}\rightarrow \texttt{[<,V,V]}\rightarrow\texttt{[=,V,-,V,V]}\rightarrow \texttt{[CALL,\#4,V]}\rightarrow \texttt{[=,V,+,V,L]}$
     \vspace*{0.1cm}
   \item[] Code fragment 2: for (int l=0; l$<$m; l++) \{ acc=acc+l; \} \\
           \hspace*{1.5in}\includegraphics[width=0.02\linewidth]{fig/Pfeil.eps}
          \vspace*{-.4cm}
   \item[]$g=\texttt{[=,V,L]}\rightarrow \texttt{[<,V,V]}\rightarrow\texttt{[=,V,+,V,V]}\rightarrow$
   $\texttt{[=,V,+,V,L]}$
    \end{itemize}
   \textbf{Hamming:}
     \begin{itemize}
     \item Substitute $\texttt{[=,V,+,V,V]}$ with $\texttt{[=,V,-,V,V]}$\\
        $g=\texttt{[=,V,L]}\rightarrow\texttt{[<,V,V]}\rightarrow\texttt{[=,V,-,V,V]}\rightarrow\texttt{[=,V,+,V,L]}$ 
     \item Substitute $\texttt{[=,V,+,V,L]}$ with $\texttt{[CALL,\#4,V]}$\\
        $g = \texttt{[=,V,L]}\rightarrow\texttt{[<,V,V]}\rightarrow\texttt{[=,V,-,V,V]}\rightarrow\texttt{[CALL,\#4,V]}$
\item[] Hamming distance $\delta_\mathit{H}(f,g)=3=2+1$ because $f$ has one more character than $g$.
     \end{itemize}
  \textbf{Levenshtein:}
  \begin{itemize}
  \item Substitute $\texttt{[=,V,+,V,V]}$ with $\texttt{[=,V,-,V,V]}$\\
          $g=\texttt{[=,V,L]}\rightarrow\texttt{[<,V,V]}\rightarrow\texttt{[=,V,-,V,V]}\rightarrow\texttt{[=,V,+,V,L]}$ 
   \item Insert $\texttt{[CALL,\#4,V]}$
   \item[] $g=\texttt{[=,V,L]}\rightarrow \texttt{[<,V,V]}\rightarrow\texttt{[=,V,-,V,V]}\rightarrow \texttt{[\mbox{CALL},\#4,V]}\rightarrow \texttt{[=,V,+,V,L]}$
     \item[] Levenshtein distance $\delta_\mathit{L}(f,g)=2$.
  \end{itemize}  
  \textbf{Needleman-Wunsch:}
  \begin{itemize}
  \item[] $f=\texttt{[=,V,L]}\rightarrow \texttt{[<,V,V]}\rightarrow \_\_\_\_\_\_
          \rightarrow \texttt{[=,V,-,V,V]}\rightarrow \texttt{[CALL,\#4,V]}$ 
            $\rightarrow \texttt{[=,V,+,V,L]}$
     \item[] $g = \texttt{[=,V,L]}\rightarrow\texttt{[<,V,V]}\rightarrow\texttt{[=,V,+,V,V]}\rightarrow \_\_\_\_\_\_
\rightarrow \_\_\_\_\_\_
 \rightarrow \texttt{[=,V,+,V,L]}$  
     \item[] Needleman-Wunsch-based distance $\delta_\mathit{NW}(f,g)=max(|f|-score(f,g),|g|-score(f,g)) = 2$, where $score(f,g) = 3$.
  \end{itemize}
\textbf{Longest Common Subsequence (LCS):}
\begin{itemize}
   \item[] $f=\texttt{[=,V,L]}\rightarrow \texttt{[\textless,V,V]}\rightarrow\texttt{[=,V,-,V,V]}\rightarrow \texttt{[CALL,\#4,V]}\rightarrow \texttt{[=,V,+,V,L]}$
    \item[] \vspace{0.1cm}$g=\texttt{[=,V,L]}\rightarrow \texttt{[\textless,V,V]}\rightarrow\texttt{[=,V,+,V,V]}\ \rightarrow \texttt{[=,V,+,V,L]}$
   \item[] $z=\texttt{[=,V,L]}\rightarrow \texttt{[\textless,V,V]}\rightarrow \texttt{[=,V,+,V,L]}$
\item[] LCS-based distance
$\delta_\mathit{LCS}(f,g)=max(|f|-|z|,|g|-|z|) = 2$.
\end{itemize}
  }
\caption{Distance metrics for calculating path similarity}
\label{fig:metrics}
\end{figure}

A number of string metrics can be borrowed from bioinformatics, where
they are usually used for DNA sequence alignment. One such metric is the
\emph{Levenshtein distance}~\cite{DBLP:journals/iandc/Levenshtein75}.
This distance measure is defined as the minimum number of mutating operations
for two given text strings, i.e., inserting, deleting, or substituting a
single character, which are needed to convert the one string into the other.
In other words, the Levenshtein distance is an extension to the previously
described Hamming distance, where text strings are not limited to having the
same length, and the insertion and deletion operations are additionally
allowed besides character substitution.

Another metric that has its origins in the alignment of DNA sequences is
the \emph{Needleman-Wunsch algorithm}~\cite{Needleman1970}. The algorithm
computes an alignment of two given text strings that is optimal with
respect to a certain scoring scheme using two steps. In the first step,
a similarity matrix of the two text strings is generated based upon the
scoring scheme, which defines weights for character matches, mismatches,
and gaps. In the second step, the optimal string alignment with maximal
total score is computed by performing a backward traversal on this
similarity matrix. \textcolor{black}{While the Levenshtein and Needleman-Wunsch-based distance
require more computational effort, they are also more permissive in case
of code clones with larger syntactical variance as in case of large-gap
clones, where matching code fragments have many and/or larger
edits similar to DNA sequence alignments~\cite{DBLP:journals/access/LiuWFWL19}.}

The application of string metrics to the comparison of paths in abstract
dominator trees is illustrated in Fig.~\ref{fig:metrics} by way of an
example. In the upper part, two code fragments are shown, conjoined with
two paths $f$ and $g$ from their abstract dominator trees. As can be seen
below, the (abstract) instructions take on the role of characters in the
visualized calculations of string metrics. In the alignment provided for the
Needleman-Wunsch algorithm, vertical lines between nodes represent matches
and horizontal lines represent inserted gaps. For the value of $score(f,g)$,
a scheme is used in which the weights for matches are set to $1$, for
mismatches to $-1$, and for gaps to $0$. 

The \emph{Longest Common Subsequence
(LCS)}
algorithm~\cite{DBLP:conf/spire/BergrothHR00,DBLP:conf/iwpc/RoyC08a} is
a well-known tool for comparing the contents of files, used for example
in the Unix \texttt{diff} utility or in the revision control system
\texttt{git}. Besides, clone detectors like NiCad are relying on the
LCS algorithm as well. In our case, the algorithm is applied to paths
instead of lines of code, such that the longest common subpath is
computed, e.g., $\texttt{[=,V,L]}\rightarrow \texttt{[<,V,V]}\rightarrow \texttt{[=,V,+,V,L]}$
for paths $f$ and $g$ in Fig.~\ref{fig:metrics}. This longest common
subpath then forms the basis for defining the metric $\delta$ for two
paths of dominator trees, which equals the maximum number of unique
nodes for both paths.

\textcolor{black}{Using the original LCS-based metric, though, can become
problematic in cases of large-gap code clones~\cite{DBLP:conf/icse/WangSWXR18},
e.g., code clones where one code fragment is embedded into the other. The
LCS metric then has problems recognizing the similarity of the
respective paths of the dominator trees of the code fragments, as the
paths differ too greatly in length. For this reason, a modified
variant of the LCS-based metric can be used, which performs an additional
similarity check in situations where two paths are not considered similar
by the original LCS-based metric and therefore analyzes whether the length
of the derived longest common subsequence is close to the length of the
respective smaller path. If this is the case, the unique nodes in the
smaller path are considered for building the LCS metric measure, rather
than those of the larger path. Technically speaking, code fragments that
consist of only one basic block are considered code clones if the
calculated LCS value corresponds to a value that is less than a
percentage of the number of nodes of the smaller code fragment that
is specified by a certain threshold.}

\textcolor{black}{This additional detection feature for subclones can also
be applied to such code fragments that do not consist of only one basic block.
For that purpose, for two subpaths $q$ and $p$, for which the length of the
longest common subsequence is close to the number of nodes of the smaller
path $p$, the LCS value is determined using the uniquely occurring nodes in
$p$. However, to avoid overly imprecise clone analysis, the examination for
subpaths only makes sense if the length of the longest common subsequence of
the two paths is nearby to the length of the smaller path $p$ and, furthermore,
a minimum length is assumed for $p$. In our later described evaluation experiments
(cf. Sect.~\ref{sec:settings}), we observed that a maximum length difference of
10\% in the number of nodes in the longest common subsequence and $p$ and a
minimum length of 10 lines for $p$ is a good practical choice for these
parameters.}

Note that there exist other string metrics as well, besides the ones
discussed above, which have also been implemented in \textcolor{black}{StoneDetector}.
Among these are, e.g., N-gram distance, cosine distance,
Jaccard similarity, etc. For the sake of brevity, we here limit ourselves
to the subset of metrics that we use in the evaluation in
Sect.~\ref{sec:evaluation}. For a more comprehensive discussion of
string metrics, we refer, e.g., to the survey~\cite{h2013survey}. 

\subsection{Fingerprints and Hashing} \label{sec:hashing}

However, using string metrics on the plain paths of (abstract) dominator
trees for path comparison is rather inefficient due to the need for node
matching, i.e., matching equal instructions, multiple times for a
single comparison of two paths. Therefore, to minimize the costs,
we conduct the node matching on fingerprints instead, which encode
the respective nodes' instructions. Such a \emph{fingerprint} is simply
the hashed value of an instruction, for which the string metrics are
adjusted accordingly. There are apparently various hashing algorithms
available for this purpose. For example, the
\emph{MD5 algorithm}~\cite{DBLP:journals/rfc/rfc1321} provides a widely
used hash function with 16-byte hash values, which is often used as a
checksum of file downloads and suffices for creating collision-free
encodings of instructions in our abstract dominator trees. The MD5
algorithm is the default hashing algorithm for fingerprints in
\textcolor{black}{StoneDetector}, and its application for encoding
the description sets of the example in Fig.~\ref{fig2} is shown in the
lower part of the same figure.

We have implemented and experimented with two other hashing algorithms
as well, including a simple 4-byte hash function based upon prime number
encoding and Locality-Sensitive Hashing. For the latter, we
observe that the MD5 hashing algorithm deliberately generates different
hash values for distinct instructions in the description sets in order
to avoid collisions. For instance, the two instructions \lstinline[columns=fixed]{i=-i*2}
and \lstinline[columns=fixed]{i=i*-2} are mapped to different hash values
even though they are syntactically and semantically very similar statements.
As a result, similar paths and description sets might not be identified as
such, and code clones might therefore be missed.

By contrast, \emph{Locality-Sensitive Hashing (LSH)~\cite{datar04}} can be
used to create identical hash values for similar elements. LSH places
elements to be mapped to hash values into buckets so that similar elements
are most likely to be stored in the same bucket. Within the algorithm, the
number of buckets is always chosen so that it is significantly smaller
compared to the number of elements. In previous work, LSH has already 
been used for code clone detection (cf.~\citet{alomart2020}).
For that purpose, the number of comparisons between code fragments
is reduced by placing similar fragments in the same bucket.
When similarity of code fragments is later analyzed, only those fragments 
in the same function have to be considered.

In StoneDetector, LSH can be used as an alternative to MD5 as well, though
at a more fine-grained level. Using LSH allows for assigning the same hash
values to similar nodes. To this end, the nodes contained in description
sets are clustered according to the length of the represented
instruction, and the similarity-preserving hash values are determined
separately for the resulting clusters. By default, StoneDetector
uses the length clusters 3 to 8. Instructions of the corresponding length
are assigned to the length clusters, whereby instructions with a length
greater than 8 are also assigned to the length cluster 8. LSH generates
200 buckets for each of the different length clusters, in which the
instructions assigned to the individual length clusters are then
organized according to their similarity.

\subsection{Comparison of Description Sets} \label{sec:comparison}

Choosing one of the metrics in combination with a hashing algorithm,
as presented in Sect.~\ref{sec:metrics} and Sect.~\ref{sec:hashing},
for the distance function $\delta$
allows us to compare description sets and subsequently assess the
similarity of code fragments.
To this end, given two code fragments with respective description
sets $D_\mathit{set}$ and $D'_\mathit{set}$, the most similar path
$q$ of $D'_{\mathit{set}}$ is identified for each path $p$ of
$D_{\mathit{set}}$, such that $q$ has a minimal value $\delta(p,q)$,
and the arithmetic mean for these values over all paths
$p\in D_{\mathit{set}}$ is computed. Formally, we can define
the distance of the two description sets
$D_\mathit{set}, D'_\mathit{set}\in \mathbb{D_\mathit{set}}$, having
$|D_{\mathit{set}}| \leq |D'_{\mathit{set}}|$ without loss of generality,
as a function $\Delta$ such that: \smallskip


{\centering
 \vspace{0.1cm}
$\Delta: \mathbb{D_\mathit{set}} \times \mathbb{D_\mathit{set}} \rightarrow \mathbb{N}$, with\\
  \vspace{0.3cm}
$\Delta(D_{\mathit{set}}, D'_{\mathit{set}}) =  \sum\limits_{p\in D_{\mathit{set}}} min(\{\delta(p,q)\,|\,q \in D'_{\mathit{set}}\}) / |D_{\mathit{set}}| $\\
}
\smallskip

We are, though, not interested in whether two description sets are
identical, rather, we want to find code fragments having similar
description sets, i.e., code clones. A threshold $\tau$
is therefore defined, limiting the maximum difference of description sets that are
considered as similar. Using $\tau$, we have
$\Delta(D_{\mathit{set}}, D'_{\mathit{set}})<\tau$ for
two description sets $D_{\mathit{set}}, D'_{\mathit{set}}\in \mathbb{D_\mathit{set}}$ iff
$D_{\mathit{set}}, D'_{\mathit{set}}$ are similar and their
corresponding code fragments constitute a code clone.

\section{Evaluation} \label{sec:evaluation}

In the evaluation, we are interested in finding answers to a number of 
research questions, in particular about the performance of our code
clone detector compared to the current state of the art. We also
present results that we believe to provide general insights into
code clone detection for Java and to help other researchers when
implementing code clone detectors, besides showing the versatility
of \textcolor{black}{StoneDetector}. 
Thus, we address the following research questions: 
\begin{itemize}
    \item \textcolor{black}{\emph{Research Question 1:} How many code clones does
          our clone detection method and tool StoneDetector
          detect, i.e., what is its recall compared to the state
          of the art, and how many of StoneDetector's reported code clones
          are real code clones, i.e., what is the tool's precision
          compared to the state of the art?}
    \item \textcolor{black}{\emph{Research Question 2:}}
          \textcolor{black}{Is the conventional code clone detector StoneDetector
          in general able to find more difficult-to-identify code
          clones, i.e., code clones with larger syntactical
          deviation?}
    \item \textcolor{black}{\emph{Research Question 3:}} Does StoneDetector scale to
          large codebases containing up to multiple million lines
          of code?
    \item \textcolor{black}{\emph{Research Question 4:} How do the used threshold
          value and the minimal size of code clone influence
          StoneDetector's performance? Which role do different
          string metrics and hashing methods play?}
    \item \textcolor{black}{\emph{Research Question 5:}} Can source code clones also
           be found in Java Bytecode using clone detectors like
           StoneDetector?
\end{itemize}

\begin{table}
\centering
\footnotesize
\scalebox{0.9}{
\begin{tabular}{l|l}
Tool & Configuration parameters \\
\hline
\emph{Deckard} & 70\% similarity, 2 token stride \\ \hline
\emph{SourcererCC} & function granularity, 70\% similarity\\ \hline
\multirow{2}{*}{\emph{NiCad 6.2}} & blind renaming, literal abstraction, \\
& 70\% similarity threshold \\ \hline
\emph{iClones} & informat single, minclone 50, minblock 15 \\ \hline
\multirow{2}{*}{\emph{CCAligner}} & 60\% threshold, 6 window size, \\
& 1 edit distance \\ \hline
\multirow{2}{*}{\emph{StoneDetector}} & 70\% similarity threshold ($\tau = 0.3$), \\
 &  metric $\delta_{\mathrm{LCS_{modified}}}$, LSH hashing \\ \hline
\emph{Oreo} & 55\% action filter, 60\% partition threshold \\ \hline
\multirow{2}{*}{\emph{NIL}} &  5-gram, 10 partitions, 10\% filter threshold, \\
&  70\% threshold \\ \hline
\multirow{2}{*}{\textcolor{black}{\emph{CloneWorks}}} & \textcolor{black}{70\% similarity threshold, aggressive configuration:} \\
& \textcolor{black}{line splitting, blind renaming, literal abstraction} 
\end{tabular}
}
\captionof{table}{Tools' configuration parameters}
\label{tbl:configuration}
\end{table}

For our evaluation, we have mainly chosen
BigCloneBench~\cite{DBLP:conf/icsm/SvajlenkoR15,DBLP:conf/icsm/SvajlenkoR16}\footnote{\url{https://github.com/jeffsvajlenko/BigCloneEval}}
as the
state-of-the-art benchmark for Java code clone detection. However,
in order to avoid an evaluation skewed to this benchmark and to
assess the performance of clone detectors for the more
difficult-to-identify code clones of Type 3 and Type 4, we also
consider the benchmarks
Google Code Jam\footnote{\url{https://zibada.guru/gcj/}},
Project CodeNet~\cite{DBLP:conf/nips/Puri0JZDZD0CDTB21}\footnote{\url{https://github.com/IBM/Project_CodeNet}},
and GPTCloneBench~\cite{DBLP:conf/icsm/AlamRARRS23}\footnote{\url{https://github.com/srlabUsask/GPTCloneBench}}.

\textcolor{black}{In the comparison of StoneDetector with other code clone
detectors for Java, a set of representative state-of-the-art
tools has been chosen, including conventional text-based,
token-based, tree and graph-based, as well as hybrid clone
detectors (cf. Sect.~\ref{sec:relatedwork}).
While we focus on conventional code clone detectors in this
paper, we have also added the Oreo hybrid tool, which 
also implements a Deep Learning approach, for the sake
of comparison.}
 Namely, we use
the following tools:
\begin{itemize}
 \item \emph{NiCad}~\cite{DBLP:conf/iwpc/RoyC08a,DBLP:conf/iwpc/CordyR11a}\footnote{\url{https://www.txl.ca/txl-nicaddownload.html}},
 \item \emph{SourcererCC}~\cite{DBLP:conf/icse/SajnaniSSRL16}\footnote{\url{https://github.com/Mondego/SourcererCC}},
 \item \emph{CCAligner}~\cite{DBLP:conf/icse/WangSWXR18}\footnote{\url{https://github.com/PCWcn/CCAligner}},
 \item \emph{iClones}~\cite{DBLP:conf/csmr/GodeK09}\footnote{Code received via  correspondence with tool's authors.},
 \item \emph{Deckard}~\cite{DBLP:conf/icse/JiangMSG07}\footnote{\url{https://github.com/skyhover/Deckard}},
 \item \emph{NIL}~\cite{DBLP:conf/sigsoft/NakagawaHK21}\footnote{\url{https://github.com/kusumotolab/NIL}},
 \item \emph{Oreo}~\cite{DBLP:conf/sigsoft/SainiFLBL18}\footnote{\url{https://github.com/Mondego/oreo-artifact/}},
 \item \textcolor{black}{\emph{CloneWorks}~\cite{DBLP:conf/icse/SvajlenkoR17a}}\footnote{\textcolor{black}{\url{https://github.com/jeffsvajlenko/CloneWorks}}}
\end{itemize}.

\begin{table}
\centering
\footnotesize
\scalebox{0.9}{
\begin{tabular}{l|c|c|c|c|c|c}
\multirow{2}{*}{Benchmark} & \multirow{2}{*}{Type 1} & \multirow{2}{*}{Type 2} & Very-Strongly & Strongly & Moderatly & \multirow{2}{*}{Type 4} \\
& & & Type 3 & Type 3 & Type 3 &  \\ \hline
\emph{BigCloneBench} & 47,146 & 4,609 & 4,163 & 16,631 & 83,444 & 8,219,320 \\ \hline
\emph{GPTCloneBench} & \multicolumn{2}{c|}{5,431} & \multicolumn{4}{c}{11,558} \\ \hline
\emph{Google Code Jam} & \multicolumn{6}{c}{10,936,986} \\ \hline
\emph{Project CodeNet} & \multicolumn{6}{c}{11,212,500}
\end{tabular}
}
\captionof{table}{Benchmarks (clone numbers vary according to chosen minimal clone length)}
\label{tbl:benchmarks}
\end{table}

In the evaluation experiments, we only consider code fragments with a minimal
length of 15 lines in the original source code if not otherwise stated.
Considering smaller code fragments in general results in a skewed comparison.
For instance, in a Java method with only 6 lines of source code, half of the 
lines are usually just taken by the method's annotations, signature, and
curly braces. However, for the curious reader, we also provide experiments
regarding different minimal sizes of code fragments (cf. Sect.~\ref{sec:settings}).
The clone detectors' configurations are
otherwise chosen in conformance with the literature; see also
Table~\ref{tbl:configuration}.
\textcolor{black}{In case of StoneDetector, the tool' configuration has
been chosen for the tool's best tradeoff of recall and precision with
acceptable runtime. When deviating from this default configuration,
we will explain it in the respective sections.} All experiments have been
performed on a Linux workstation with an Intel Core i7-10700K processor
and 32 GB RAM.

Note that we provide an online version of StoneDetector, the tool's
source code, a runnable container image, and supplementary material,
which we here omit due to space
constraints\footnote{\url{https://stonedetector.fmi.uni-jena.de/}}~\cite{DBLP:conf/iwsc/SchaferAH23,burock,ostryanin,schaefer}.

\subsection{Benchmarks} \label{sec:benchmark}


The benchmark \emph{BigCloneBench}~\cite{DBLP:conf/icsm/SvajlenkoR15} is the state-of-the-art
benchmark for code clone detection in Java source code. It contains 8,379,922 code clone pairs
over 55,499 Java source files with approximately 15.4 million lines of code. The benchmark has been
assembled from real-world software projects of the \emph{IJaDataset 2.0}~\cite{ijadataset}. To this
end, methods in the IJaDataset have been heuristically classified and manually validated
according to a set of 43 different functionalities, e.g., bubble sort or database update and
rollback. If two methods are then classified as having the same functionality, they are considered a
code clone pair. The resulting code clones are examined and classified according to their
clone type, i.e., Type 1, Type 2, Type 3, and Type 4 code clones,
as introduced in Sect.~\ref{sec:introduction}. The latter two are further differentiated
into:
\begin{itemize}
\item \emph{Very-Strongly Type 3} with 90-100\% syntactical similarity after normalization
\item \emph{Strongly Type 3} with 70-90\% syntactical similarity after normalization
\item \emph{Moderately Type 3} with 50-70\% syntactical similarity after normalization
\item \emph{Weakly Type 3/Type 4} with less than 50\% syntactical similarity after normalization
\end{itemize}
Note that the benchmark BigCloneBench is quite skewed towards harder-to-find Type 4
code clones, as can be seen in Table~\ref{tbl:benchmarks}.

\textcolor{black}{The usage of BigCloneBench as dataset for harder-to-detect code clones of Type 3 and Type 4 have been controversially discussed in recent years\cite{DBLP:conf/issre/LiGZ23,DBLP:journals/corr/abs-2505-04311}.
In particular, its naive usage as dataset for Deep learning clone detectors can compromise training and evaluation~\cite{DBLP:conf/iwsc/SchaferAH22,DBLP:conf/iwsc/KrinkeR22}. The benchmark suffers from several issues, which are caused by its
construction. For example, the benchmark is not only skewed towards Type 4 code clones but also towards certain functionalities, where, e.g., the Copy file functionality holds half of all code clones~\cite{DBLP:journals/corr/abs-2505-04311}. Besides, its ground
truth quality, i.e., true code clones labels, is disputed for Type 4 code clones~\cite{DBLP:conf/issre/LiGZ23,DBLP:journals/corr/abs-2505-04311}. However, note that these issues do not impair its usage for evaluating clone detectors on Type-1, Type-2, and Type-3 clones. However, in order to underpin our findings for harder-to-detect code clones of Type 3 and Type 4, we augment our evaluation with other state-of-the-art benchmarks for code clone detection.}

A benchmark including in particular harder-to-detect code clones has been defined
with \emph{GPTCloneBench}~\cite{DBLP:conf/icsm/AlamRARRS23}. The authors of the benchmark
used the code clones of the \emph{SemanticBench}~\cite{DBLP:conf/iwsc/Al-OmariRC20}
benchmark in prompts for generating additional
code clones with the GPT-3 large language model. Manually validating the generated
code clones afterward resulted in a set of 37,149 code clones, of which 16,989 are 
clone pairs with Java source code that we can use in our experiments.
In spite of the benchmark's authors intentions, we use the 5,431 near-miss code
clones of Type 1 and Type 2 as well as the 11,558 Type 3 and Type 4 code clones
in the experiments. 
\emph{Google Code Jam} was an
international programming contest where
contestents were solving programming tasks. The posted solutions are
publicly available, comprising 3.5 million
code fragments in different programming languages for 716 programming tasks.
Two different solutions to the same programming task can apparently be
considered to share the same computations and thus to constitute a code clone.
The thus defined code clones can vary across the different types of code clones.
We consider solutions for the year 2022, and, as only part of the solutions are Java
code, we are able thus to use 10,936,986 code clone pairs in our experiments.
\textcolor{black}{Note that Google Code Jam also suffers from being biased towards certain
dominant functionalities, i.e., programming tasks~\cite{DBLP:conf/issre/LiGZ23}.}
The last benchmark also originates from programming contests.
\emph{Project CodeNet}~\cite{DBLP:conf/nips/Puri0JZDZD0CDTB21} assembles 14 million
code fragments and more than
500 million lines of code with solutions to 4053 programming tasks, written
in 55 programming languages. Different solutions to the same programming task
are again considered code clones, however, near-miss code
clones, i.e., code fragments sharing above 90\% syntactical similarity,
are omitted in this benchmark~\cite{DBLP:conf/nips/Puri0JZDZD0CDTB21}.
Around 5\% of the code fragments, or 25 million lines of code, in
Project CodeNet are Java code, which results in 11,212,500 code clone
pairs used in our experiments.
\textcolor{black}{Apparently, in case of Google Code Jam and Project CodeNet, code clone
samples from programming contests may not be representative of production
code~\cite{DBLP:conf/issre/LiGZ23}. We though believe that using a wide
range of benchmarks allows for a generalizable interpretation of our
evaluation results, even if we are aware of the benchmarks' issues.} 
Note that all three benchmarks,
in contrast to BigCloneBench, do not distinguish between the
different clone types.

\begin{table}
  \centering
 \footnotesize
\scalebox{0.675}{
    \begin{tabular}{l|c| c | c  | c|c | c |c|c|c}
	    &\emph{Deckard}&\emph{SourcererCC}&\emph{NiCad}&\emph{iClones}&\emph{CCAligner}&\emph{StoneDetector}&\emph{NIL}& \textcolor{black}{\emph{CloneWorks}}&\emph{Oreo} \\ \hline
	    Recall    & 18,318 &  20,827 &  20,829 & 20,842 & 20,831 & 20,828 &  \textbf{20,844} &  \textcolor{black}{20,828} & \textbf{20,844} \\
	    T 1  & 87.8\% &  99.9\%  & 99.9\%  & 99.9\%  & 99.9\%  & 99.9\%  & \textbf{100\%}  &  \textcolor{black}{99.9\%} &  \textbf{100\%} \\
  \hline
	    Recall  & 3,317 & 3,444 & 3,460 & 3,453 &  3,456 & \textbf{3,475} &  3,460 &  \textcolor{black}{3,460}  &  \textbf{3,475}  \\
	    T 2 & 95.4\% & 99.1\% & 99.5\% & 99.3\% & 99.4\% & \textbf{99.9\%} & 99.5\% &   \textcolor{black}{99.5\%} & \textbf{99.9\%} \\
  \hline
	    Recall	& 2,139 &  3,209  & 3,243  & 3,253 & 3,218 & 3,209 & 3,271 &   \textcolor{black}{3,240} &  \textbf{3,283}  \\
	    VST 3 & 65.0\% & 97.6\% & 98.6\% & 98.9\% & 97.8\% & 97.6\% & 99.5\% &   \textcolor{black}{98.5\%} & \textbf{99.8\%} \\
  \hline
	    Recall  & 1,806  & 5,784  & 6,246 & 6,531 & 5,943 & 6,808  & 6,549 &  \textcolor{black}{\textbf{7,073}} & 6,584 \\
	    ST 3 & 24.4\% & 78.1\% & 84.3\% & 88.2\% & 80.2\% & 92.0\%  & 88.5\% &  \textcolor{black}{\textbf{95.5\%}} & 88.9\% \\
  \hline
	    Recall & 2,311  & 1,462  & 169 & 4,051 & 1,784 & \textbf{6,324} & 3,499 &  \textcolor{black}{1,305} & 5,187  \\
	    MT 3 & 8.6\% & 5.5\% & 0.6\% & 15.1\% & 6.6\% & \textbf{23.6\%} & 13.1\% &  \textcolor{black}{4.8\%} & 19.4\% \\
    \hline
	    Recall  & \textbf{57,063} & 250 & 7 & 6,554 & 2,133 & 6,061  & 3,821 &  \textcolor{black}{19} & 6,821  \\
	    T 4 & \textbf{1.3\%}  & 0.0\% & 0.0\%  & 0.1\%  & 0.0\%  & 0.1\%   & 0.1\%  &   \textcolor{black}{0.0\%} & 0.1\%   \\
\hline \hline
	    \#Clones  & 13,239,074  & 804,315  & 754,022  & 987,659  & 1,244,160  & 1,300,458  & 1,274,852 &  \textcolor{black}{915,434} & 1,015,425 \\ \hline

	    Precision  & 46\%  & \textbf{100\%}  & \textbf{100\%}  & 97\%  & 98\%  & 99\%  & 99\%  &   \textcolor{black}{\textbf{100\%}} & 99\%
    \end{tabular}
}
  \captionof{table}{Recall and precision of clone detectors on Java source code on BigCloneBench
(T 1 - Type 1, T 2 - Type 2, VST 3 - Very-Strongly Type 3, ST 3 - Strongly Type 3, MT 3 - Moderately Type 3,
 T 4 - Type 4, \#Clones - total number of detected clones)} 
  \label{tbl:results}
\end{table}

\subsection{\textcolor{black}{Research Question 1: Recall and Precision}} \label{sec:bigclonebench}

In order to address the first research question about how many
code clones in Java source code our clone detection method is able to find, we ran
StoneDetector on BigCloneBench via BigCloneEval~\cite{DBLP:conf/icsm/SvajlenkoR16}.
For a comparison with the state of the art, we also ran seven other
clone detectors on the same benchmark. The resulting number of identified known code 
clones, i.e., \emph{recall values}, for the tools Deckard, SourcererCC,
NiCad, iClones, CCAligner, StoneDetector, NIL, \textcolor{black}{CloneWorks},
and Oreo are shown in
Table~\ref{tbl:results}, as measured by BigCloneEval.
Since these numbers reflect only the annotated code clones in the benchmark
(cf. Sect.~\ref{sec:benchmark}), we additionally list the overall
number of detected clones. As noted before, we only consider clones
with a minimum size of 15 lines of code. Duplicate clones, as reported
by some tools, were removed. Recall values are given in absolute
and relative numbers. 

As can be seen in the table, all the code clone detectors, with the exception
of Deckard, perform well on exact and near-miss code clones, i.e.,
code clones of Type 1, Type 2, and Very-Strongly Type 3. The respective recall
values are located in the interval 97-100\% and only for Deckard do we observe
a worse recall value of 65\%. These numbers in general reflect the results of prior
research~\cite{DBLP:conf/icse/SajnaniSSRL16}. Thus, we can state that
StoneDetector performs competitively with the state of the art for exact and
near-miss code clones. \textcolor{black}{Note though that in case
of Type 1 clones, i.e., exact duplicates of code fragments that only differ
in layout, comments, or formatting (cf. Sect~\ref{sec:introduction}),
StoneDetector misses certain code clones.
Closer inspection revealed that these cases can be attributed to a low number
of code samples from the benchmark which can not be properly processed or
parsed using the Spoon analysis framework.}

As expected, beginning with code clones of Strongly Type 3, recall values
are declining due to the larger syntactical differences of code fragments.
We can also observe a difference between near-miss code clone detectors,
e.g., NiCad\textcolor{black}{, CloneWorks,} and SourcererCC, and tools addressing harder-to-find clones,
i.e., CCAligner, iClones, Oreo, StoneDetector, and NIL. StoneDetector
achieves the best recall values for code clones of Strongly Type 3 and
Moderately Type 3, but is overtaken by the tools Deckard, iClones, and
Oreo for code clones of Type 4. However, the high number of Type 4 clones
Deckard is finding has to be taken with care, considering its poor
precision as described in the following section and also reported in the
literature~\cite{DBLP:conf/icse/SajnaniSSRL16,DBLP:conf/icse/WangSWXR18}.


In addition to a high number of identified code clones, a code clone detector
shall also avoid reporting on wrong code clones. Thus, we have to measure
the clone detector's \emph{precision}, i.e., how many of the reported code clones
are actual code clones. Unfortunately, answering \textcolor{black}{this question}
is challenging, since BigCloneBench/BigCloneEval only contains information
about a subset of all possible code clones in the benchmark. Due to the
benchmark's construction, only code clones of the considered 43 functionalities
are known~\cite{DBLP:conf/icsm/SvajlenkoR15}. We have therefore performed a
manual evaluation on random samples of the tool's reported code clones, as
is also usually conducted in the respective research
literature~\cite{DBLP:conf/icse/WangSWXR18,DBLP:conf/sigsoft/SainiFLBL18,DBLP:conf/icse/SajnaniSSRL16}.

To this end, we randomly selected samples of 400 clone pairs of the overall code
reported clones for each clone detector. The clone pairs were blindly, i.e.,
without revealing the respective tool, given to the three authors of this paper,
who acted as judges. Each judge individually assessed the given clone pairs
as actual or wrong clone pairs (true/false positive). Ambiguous assessments
between the three judges were discussed and resolved afterward.

The resulting precision estimates for the clone detectors are also shown
in Table~\ref{tbl:results}. As already mentioned, Deckard achieves subpar
precision, i.e., only 46\% of Deckard's reported code clones are assessed
as such by the three judges. This observation has already been reported for
prior evaluations~\cite{DBLP:conf/icse/WangSWXR18,DBLP:conf/sigsoft/SainiFLBL18}.
The other tools, including StoneDetector, perform at high precision levels
between 97-100\%, while near-miss clone detectors NiCad and SourcererCC achieve
the best precision. Compared to prior evaluations, we find in general better
precision numbers. This can, though, be attributed to the chosen minimum size
of 15 lines of code clones, compared to the 6 lines used in other work.
Increased minimum clone size supports the precision of clone detectors,
as has already been reported by~\citet{DBLP:conf/icse/WangSWXR18}.
\bigskip \\
\noindent\fbox{
    \parbox{.95\textwidth}{
        \textcolor{black}{When compared with state-of-the-art conventional clone detectors
        for Java source code, StoneDetector detects code clones with
        competitive recall and finds more code clones with 30\% to
        50\% syntactical variance. At the same time, StoneDetector detects
        code clones in Java source code, similar to
        other state-of-the-art clone detectors, at very high precision.}
    }
} \bigskip

\subsubsection{\textcolor{black}{Qualitative Analysis}}
\label{sec:examination}

We conducted further studies on the characteristics of the code
clones found by StoneDetector and the other clone detectors. To this
end, we first made pairwise comparisons of the detected code clones for
StoneDetector and some of the other tools, respectively, and looked into
their common and exclusively identified code clones. On the one hand,
we were thereby able to observe a large fraction of commonly identified
code clones. This mostly points to the subset of exact and near-miss
clones as identified by all clone detectors (cf. also
Table~\ref{tbl:results}). \textcolor{black}{On the other hand, we also found a
significant portion of code clones with larger syntactical
variance, in particular structural code clones, i.e, code clones
which implement the same control flow using different syntactical
constructs, and subclones, i.e., code clones where the control flow
of one code fragment is embedded into the other fragment’s control
flow~\cite{DBLP:conf/icsm/AmmeHS21}. Code clones of these types are
often exclusively identified by StoneDetector and not by the other tools.}

\textcolor{black}{When analyzing clone pairs that are detected by
StoneDetector but not by the other tools, we created random samples
of found code clones on BigCloneBench, specifically for comparing
StoneDetector with NiCad, Oreo, CCAligner, and CloneWorks, as
representatives of near-miss clone detectors, tools focusing on
large-gap clones, and Deep Learning approaches. For this purpose,
for each tool $A$, we generated a random sample of 100 clones
for the set of code clones identified by StoneDetector but not
by $A$ ($StoneDetector\setminus A$) and for the set of
code clones detected by $A$ but not by StoneDetector
($A\setminus StoneDetector$), respectively, and manually
inspected those samples. Finally, we also created a random
sample of 100 code clones which were only detected by
StoneDetector and not by any other tool considered in
our evaluation, though excluding Deckard due to its low precision
(i.e., NiCad, SourcererCC, CCAligner, iClones, NIL, Oreo, and
CloneWorks).}

\paragraph{\textcolor{black}{StoneDetector vs NiCad}}

\textcolor{black}{The closer examination of 100 random clone pairs
that were detected by StoneDetector but not by NiCad revealed
that StoneDetector finds more clones with greater syntactic variance
and in particular,  structural clones (55) and subclones (35).
The remaining clones were classified as of other types (4) and
false positives (6). Analysis of 100 clones found only by NiCad
revealed that these were mainly due to NiCad's pretty printing,
which often leads to an increase in code size and thus higher
permissiveness for code clones, and due to different path sizes
in the  dominator trees in StoneDetector and token sequences in
NiCad.}

\paragraph{\textcolor{black}{StoneDetector vs CloneWorks}}

\textcolor{black}{In CloneWorks, clone detection is based on a
modified Jaccard metric.
This metric represents code fragments as sets of contained tokens
and calculates their minimum rate of overlapping tokens~\cite{DBLP:conf/icse/SvajlenkoR17a}.
Similar to Oreo, CloneWorks thus does not relies on the order of
code fragment's instructions for clone detection. As a result, CloneWorks
can also detect simple structural clones and subclones, if the lengths
of the code fragments do not differ too much. Our examination of clone pairs
found exclusively by StoneDetector showed that more structural clones (28)
and subclones (64) were detected. A closer inspection revealed that the
clones often differed greatly in the number of lines of code. Further
clones found exclusively by StoneDetector were classified as of other
clone types (6) and false positives (2). Clone pairs found exclusively
by CloneWorks can be attributed to the use of token sets. Structural
clones (24) and subclones (49) were consequently also found by CloneWorks
if the order of the instructions in the clone pairs differs.
StoneDetector cannot detect such clones because its used LCS method
requires a sequence of instructions. For example, code fragments with
a high number of identical instructions occurring at different
code locations cause issues for LCS and are often not recognized as
clones even though they contain the very same instructions. Other code
clones found exclusively by CloneWorks were such clones, in which
instructions appeared in short form in one code fragment and in full
form in the other (7), e.g., as in case of a foreach loop and a
normal for loop over iterators. In contrast to StoneDetector, which
conceives the latter loop using four sub-expressions, CloneWorks treats
both loops as one expression in its aggressive version. Remaining
code clones found exclusively by CloneWorks were again of others
clone types (17) and false positives (3).}

\paragraph{\textcolor{black}{StoneDetector vs CCAligner}}

\textcolor{black}{Our manual inspection of code clones found
by StoneDetector but not by CCAligner revealed that CCAligner
misses many subclones, which are detected by StoneDetector
(52), which can be attributed to its code window approach
failing in case of scattered code modifications. Furthermore,
StoneDetector finds structural code clones based on its applied
control flow representation of dominator trees, which are missed
by CCAligner (21). Remaining code clones were classified as of
other clone types (18) and false positives (9). In contrast,
CCAligner's code window approach allowed for finding subclones
with many concentrated code modifications which are missed by
StoneDetector (26). CCAligner, in addition, considers a function
inside an anonymous class and the surrounding code as potential
code clones, which is not the case for StoneDetector (31). The
remaining code clones exclusively detected by CCAligner were
classified as of other types (33) and false positives(10).}

\paragraph{\textcolor{black}{StoneDetector vs Oreo}}

\textcolor{black}{Oreo uses functional metrics for clone detection,
which determine, for example, the number of field accesses, array
accesses, loops, etc., in code snippets. Clone detection in Oreo
does not depend on the order of the instructions, but instead
analyzes code properties by means of these functional metrics.
In the manual review of the random samples, we found many
structural clones (49) and subclones (14) that were overlooked
by Oreo but detected by StoneDetector. These were often code
clones that were very different in structure, which StoneDetector
recognizes, but for which Oreo derives metric measures that are
too distant from each other. The remaining clones found exclusively
by StoneDetector were classified as of other clone types (31) and
false positives (6). On the other hand, most of the code clones
found exclusively by Oreo were structural clones (52) that often
contained subpaths in internal substructures which were disordered
or of very different lengths. It was also noticeable that Oreo
thereof identified subclones (25) consisting of basic blocks,
although they have very different lengths or their instructions
were of different order, which is not a problem for Oreo due to its
usage of functional metrics but pose an issues for the string
metric comparison in StoneDetector. Remaining clones found
exclusively by Oreo were assigned as of other types (21) or as
false positives (2).}

\paragraph{\textcolor{black}{StoneDetector vs All}}

\textcolor{black}{Finally, we also looked into the set of code clones
in BigCloneBench which were exclusively identified by StoneDetector
but not by the other clone detectors considered in our evaluation.
From the total of 1,300,458 code clones identified by StoneDetector
(cf. Table~\ref{tbl:results}), a fraction of 235,982 clones were not
detected by any of the other tools, i.e., almost each fifth code
clone (18\%). A closer examination of 100 randomly selected clones
of this set again showed that StoneDetector finds significantly
more subclones (49) and structural clones (35), which
often differed greatly in size. Additional clones were
classified as other (14) and false (2).}

\subsection{\textcolor{black}{Research Question 2: Generalizability}}
\label{sec:otherbenchmarks}

Our second research question addresses the capability of clone detectors to
identify code clones in source code with larger syntactical variability, in particular
code clones of Strongly Type 3 and beyond. Besides, we are also interested
in analyzing the clone detectors' performance on benchmarks other than
BigCloneBench in order to prevent the evaluation from being biased toward this
particular benchmark, \textcolor{black}{also considering the benchmark's various
issues as mentioned above.} We therefore ran the clone detectors SourcererCC,
NiCad, CCAligner, StoneDetector, NIL, \textcolor{black}{CloneWorks,} and Oreo on the three benchmarks
GPTCloneBench, Google Code Jam, and Project CodeNet, respectively, and
report the observed recall values in Table~\ref{tbl:other_benchmarks}. We omitted
Deckard and iClones due to their lower precision compared to the other
tools. We again limit clones to those with a minimal size of 15 lines,
use the same configurations, and depict recall in absolute and relative numbers.  

\begin{table}
  \centering
 \footnotesize
\scalebox{0.8}{
    \begin{tabular}{l| c  | c|c | c |c|c|c}
 Benchmark &\emph{SourcererCC}&\emph{NiCad}&\emph{CCAligner}&\emph{StoneDetector}&\emph{NIL}&\emph{Oreo}&\textcolor{black}{\emph{CloneWorks}} \\ \hline
\emph{GPTClone}    & 4,577 & 3,646 & 3,452 & 4,068 & \textbf{4,681} & 4,366 & \textcolor{black}{3,987} \\
\emph{Bench}  & 26.9\%  & 21.5\%  & 20.3\%  & 23.9\%  & \textbf{27.6\%}  & 25.7\% & \textcolor{black}{23.4\%} \\
  \hline
      \emph{Google Code}  & 101,843 & 54,363 & 102,129 & \textbf{362,989} & 212,597 & 115,879 & \textcolor{black}{91,794} \\
   \emph{Jam}& 0.9\% & 0.5\% & 0.9\% & \textbf{3.3\%} & 1.9\% & 1.1\% & \textcolor{black}{0.8\%} \\
  \hline
      \emph{Project}	& 104,790  & 121,271 & 145,218 & \textbf{713,942} & 646,232 & 537,322 & \textcolor{black}{266,888}  \\
    \emph{CodeNet}  & 0.9\% & 1.1\% & 1.4\% & \textbf{6.4\%} & 5.8\% & 4.8\% & \textcolor{black}{2.4\%}
    \end{tabular}
}
  \captionof{table}{Recall of clone detectors on GPTCloneBench, Google Code Jam, Project CodeNet} 
  \label{tbl:other_benchmarks}
\end{table}

As shown in the table, recall values are in general at the lower end
of the spectrum, which is, however, expected due to the nature of the benchmarks (cf.
Sect.~\ref{sec:benchmark}). In particular, the benchmarks Google Code Jam
and Project CodeNet contain code clones, which originate from
heterologously developed code and therefore usually include semantic code clones
with differing syntax but similar computations. Thus, the recall values
for all tools do not exceed 10\% for both benchmarks. We can as well
observe a distinction between the near-miss clone detectors NiCad, \textcolor{black}{CloneWorks,}
and SourcererCC and tools addressing more difficult code clones, i.e., CCAligner, NIL, 
Oreo, and StoneDetector, where the latter ones perform better. While the
contrast is not so apparent on the benchmark GPTCloneBench, due to its inclusion
of Type 1/2 clones, it becomes obvious for the other two benchmarks.
We also observe StoneDetector to perform the best on these
two benchmarks.
\bigskip \\
\noindent\fbox{
    \parbox{.95\textwidth}{
        StoneDetector in particular excels at finding source code clones with larger
        syntactical variance when compared to other conventional state-of-the-art
        clone detectors on different benchmarks.
    }
} \bigskip

\begin{table}
\centering
\footnotesize
\scalebox{0.95}{%
\begin{tabular}{l|c|c|c|c}
	Tool & 1 million LOC & 10 million LOC & 100 million LOC & \textcolor{black}{328 million LOC} \\ \hline
	\emph{NiCad} & 57s & 18.49min & 2.45h & \textcolor{black}{-} \\ \hline
	\emph{CCAligner} & 18s & 6.58min& -&  \textcolor{black}{-} \\ \hline
	\emph{SourcererCC} & 4.17min & 28.09min & 46.52h &  \textcolor{black}{-} \\ \hline
	\emph{iClones} & 24s & 7.26min & -&  \textcolor{black}{-} \\\hline
	\emph{StoneDetector} & 24s & 6.17min & 7.13h &  \textcolor{black}{79h} \\ \hline
	\textcolor{black}{\emph{StoneDetector}} & \multirow{2}{*}{\textcolor{black}{24s}} & \multirow{2}{*}{\textcolor{black}{2.48min}} & \multirow{2}{*}{\textcolor{black}{1.11h}} & \multirow{2}{*}{\textcolor{black}{12.32h}} \\
        \textcolor{black}{\emph{(Hamming)}} &&&& \\ \hline
	\emph{Oreo} & 2.42min & 9.54min & 7.10h &  \textcolor{black}{23.53h} \\ \hline
        \textcolor{black}{\emph{NIL}} & \textcolor{black}{\textbf{2s}} & \textcolor{black}{\textbf{17s}} & \textcolor{black}{\textbf{10.21min}} & \textcolor{black}{\textbf{1.44h}} \\ \hline
	 \textcolor{black}{\emph{CloneWorks}} &  \textcolor{black}{10s} &  \textcolor{black}{1.46min} &  \textcolor{black}{25.35min} &  \textcolor{black}{4.07h}
\end{tabular}
}
\captionof{table}{Scalability experiments on IJaDataset 2.0}
\label{tbl:runtime}
\end{table}

\subsection{\textcolor{black}{Research Question 3: Scalability}} \label{sec:scalability}

\emph{Scalability} is another important requirement for clone detectors, i.e.,
such tools need to be able to process large quantities of code in
reasonable time. In order to evaluate StoneDetector's capabilities of
analyzing large fractions of code, we use a setup described by Sajnani
et al.~\cite{DBLP:conf/icse/SajnaniSSRL16}. To this end, random subsets
of files from the IJaDataset 2.0~\cite{ijadataset} are generated,
comprising 1, 10, and 100 million lines of code, respectively. We ran
StoneDetector and other clone detectors on these subsets
and report their runtime in Table~\ref{tbl:runtime}. As can be seen,
Deckard has been omitted, since Deckard's memory consumption exceeds
available resources, as has also been noted by~\citet{DBLP:conf/icse/SajnaniSSRL16}. Furthermore,
we observe that CCAligner as well as iClones are not able to scale to
100 million lines of code (cf. also~\cite{DBLP:conf/icse/WangSWXR18}).
Also, contradicting previous results in~\cite{DBLP:conf/icse/SajnaniSSRL16},
the clone detector NiCad in its recent version NiCad v6.2 scales to
more than 100 million lines of code and also provides the lowest runtime.
We also see that StoneDetector keeps up well with the other state-of-the-art
tools and analyzes 100 million lines of code in 7 hours and 13 minutes and
10 million lines of code in under 7 minutes. Using a different string metric
as in the default configuration allows StoneDetector to run even faster
but also partially sacrifices recall and precision
(cf. Sect.~\ref{sec:settings}).

Due to these encouraging results, we performed another scalability
check using StoneDetector \textcolor{black}{and the other clone detectors}
on the entire IJaDataset 2.0~\cite{ijadataset}, comprising approx. 2.5
million Java source files with 328 million lines of code.
\textcolor{black}{The respective runtime results are as well shown in
Fig.~\ref{tbl:runtime}. As can be observed, only four of the tools
have been able to analyze the whole dataset. While CCAligner and iClones
already did not scale for 100 million lines of code, NiCad failed
for 328 million lines of code due to an internal threshold of input
size and SourcererCC was cut off after running for a whole week.}
\textcolor{black}{Overall, NIL and CloneWorks provide the best runtime,
with NIL only requiring less than 2 hours for analyzing the whole
IJaDataset}. StoneDetector has been able to analyze this very large dataset
in 79 hours. \textcolor{black}{Due to its focus on finding harder-to-detect
code clones, StoneDetector thus trades higher analysis time for recall. 
However, note that we report on a faster version of StoneDetector when
the Hamming distance is used as a string metric with MD5 hashing, as
described in Sect.~\ref{sec:settings}, which makes it a fast near-miss
clone detector. We have included the respective runtime for this
version of StoneDetector in Fig.~\ref{tbl:runtime} as well. As can be
seen, while not as fast as NIL or CloneWorks, StoneDetector then can
analyze 328 million lines of code in 12 and a half hours. Like discussed
in Sect.~\ref{sec:settings}, this version of StoneDetector though
yields significantly poorer recall values.}
\bigskip \\
\noindent\fbox{
    \parbox{.95\textwidth}{
        \textcolor{black}{StoneDetector scales to large codebases comprising
        multiple 100 million lines of code. Using the Hamming distance instead
        of the modified LCS metric makes StoneDetector behave like
        a fast near-miss clone detector.}       
    }
} \bigskip

\subsection{\textcolor{black}{Research Question 4: Tool Configuration}}
\label{sec:settings}

\begin{table}
  \centering
 \footnotesize
\scalebox{0.885}{
    \begin{tabular}{l|c | c|c | c |c|c|c}
	    &\emph{SourcererCC}&\emph{NiCad}&\emph{CCAligner}&\emph{StoneDetector}&\emph{NIL}& \textcolor{black}{\emph{CloneWorks}}&\emph{Oreo} \\ \hline
	    Recall    &  35,843 & \textbf{35,847} & 35,744 & 35,843 & 35,783 &  \textcolor{black}{35,844} & 35,480 \\
	    T 1  & 99.9\%  & \textbf{99.9\%}  & 99.6\%  & 99.9\%  & 99.8\%  &  \textcolor{black}{99.9\%} & 98.9\% \\
  \hline
	    Recall  & 4,432 & 4,550 & 4,530 & \textbf{4,579} & 4,421 &  \textcolor{black}{4,550} & 4,506 \\
	    T 2 & 96.7\% & 99.3\% & 98.8\% & \textbf{99.9\%} & 96.4\% &  \textcolor{black}{99.3\%} & 98.3\% \\
  \hline
	    Recall	 & 3,814  & 4,098 & 4,055 & \textbf{4,140} & 3,892 &  \textcolor{black}{4,094} & 3,893  \\
	    VST 3 & 91.6\% & 99.2\% & 97.4\% & \textbf{99.4\%} & 93.5\% &  \textcolor{black}{98.3\%} & 93.5\% \\
  \hline
	    Recall  & 9,026 & 10,680 & 10,534 & 13,008 & 10,083 &  \textcolor{black}{\textbf{14,165}} & 12,625 \\
	    ST 3 & 59.2\% & 70.0\% & 69.1\% & 85.3\%  & 66.1\% &  \textcolor{black}{\textbf{92.9\%}} & 82.8\% \\
  \hline
	    Recall  & 3,935 & 476 & 8,099 & 14,537 & 8,548 &  \textcolor{black}{2,782} & \textbf{22,965}  \\
	    MT 3 & 4.8\% & 0.6\% & 10.0\% & 17.9\% & 10.5\% &  \textcolor{black}{3.4\%} & \textbf{28.2\%} \\
    \hline
	    Recall  & 1,690 & 11 & 12,673 & 11,612 & 14,500 &  \textcolor{black}{46} & \textbf{52,839} \\
	    T 4  & 0.0\%  & 0.0\%  & 0.2\%  & 0.1\%   & 0.2\%  &  \textcolor{black}{0.0\%} & \textbf{0.7\%} 
    \end{tabular}
}
  \captionof{table}{Recall of clone detectors on Java source code on BigCloneBench with a minimal code clone size of 6 lines of code
(T 1 - Type 1, T 2 - Type 2, VST 3 - Very-Strongly Type 3, ST 3 - Strongly Type 3, MT 3 - Moderately Type 3,
 T 4 - Type 4)} 
  \label{tbl:otherresults}
\end{table}

Clone detection using StoneDetector should result in the highest possible recall with acceptable precision and runtime.
In order to achieve this objective, certain parameters of our clone detection method have to be chosen. Thus, our fourth
research question addresses experiments on optimal configuration parameters in StoneDetector, in particular
considering the threshold values for syntactical similarity of code clones, their minimum size, 
\textcolor{black}{the used similarity measure, i.e., string metric, and hashing algorithm for comparing and encoding code
fragments, respectively.}

\begin{figure}[t]
  \begin{center}
    \includegraphics[scale=0.25]{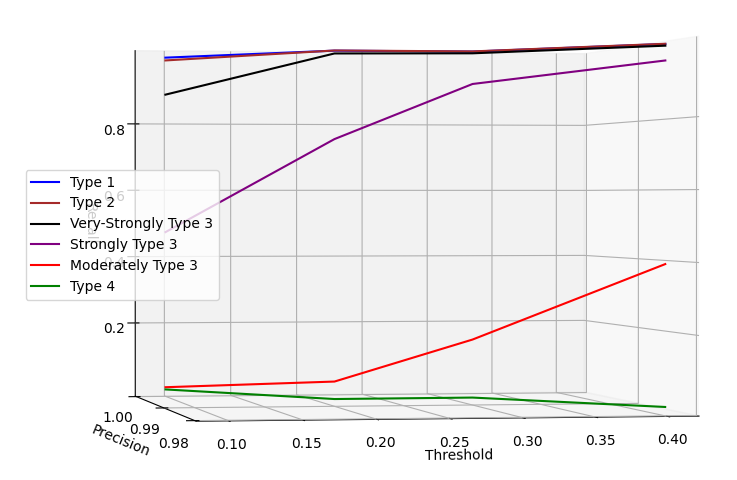} 
    \includegraphics[scale=0.22]{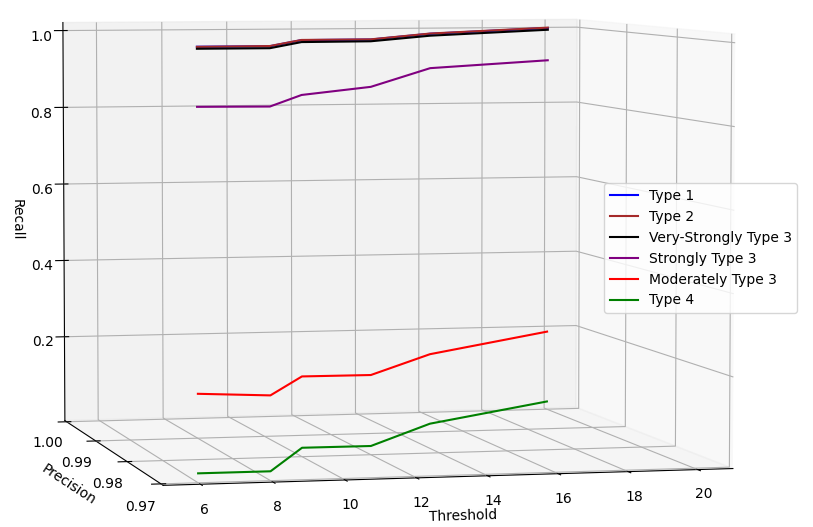} 
  \end{center}
  \caption{Additional experiments on parameter settings, considering StoneDetector's
           used distance value $\tau$ (left) and
           admitted minimal size of code clones (right)}
  \label{fig:experiments}
\end{figure}

In the experiments presented so far, we have chosen 15 lines as the \emph{minimum size of code clone fragments}.
In order to justify our previous claim that smaller code fragments in general result in a skewed
comparison, we performed additional experiments using different minimum sizes of code clones on BigCloneBench.
In Table~\ref{tbl:otherresults}, we report recall values when a minimum size of 6 lines of code is used
instead of 15 lines, since 6 lines of code are often used in the literature as well. Note that we have
again omitted the tools Deckard and iClones due to their lower precision. As expected, the
shown recall values are in general higher compared to those in Table~\ref{tbl:results}. This can well be
explained by Java methods with only 6 lines of source code, where half of the lines are just taken by the
method's annotations, signature, and curly braces. We, though, do not observe substantial changes in the tools'
comparisons, besides Oreo now outperforming StoneDetector in the case of Moderately Type 3 code clones.
Furthermore, we also considered other minimum sizes in the range from 6 to 20 lines of code, though
only for our tool StoneDetector. Due to
space constraints, instead of providing the respective numbers, we have plotted the resulting precision
and recall on BigCloneBench in Fig.~\ref{fig:experiments}. 
\textcolor{black}{Note that we provide
the detailed results as supplementary information online\footnote{\url{https://stonedetector.fmi.uni-jena.de/}}.}
As can be seen, comparable
recall values have been obtained, while precision values range from 96\% to 100\%, with increasing
values for increasing minimum sizes of code clones. We can thus confirm our stated
conclusion that increased minimum clone size supports the precision of clone
detectors~\cite{DBLP:conf/icse/WangSWXR18}.

A second parameter we are interested in is the \emph{similarity threshold}, which is used in
our method to define whether two processed code fragments are considered code clones
or not (cf. also Sect.~\ref{sec:comparison}). In the experiments conducted so far,
this threshold value has been set to 70\%, i.e., distance value $\tau=0.3$, which on the
one-hand side is comparable to the threshold values as used in other clone detectors
and on the other-hand side, in our experience, provides an optimal tradeoff of high
recall and precision. In order to confirm this optimal tradeoff, we conduct
additional experiments using StoneDetector on BigCloneBench using
differing threshold values in terms of the distance value $\tau$ ranging from
$\tau=0.1$ to $\tau=0.4$. The respective distance values are again plotted in 
conjunction with the resulting recall and precision in Fig.~\ref{fig:experiments}
(other configuration parameters equal those in Table~\ref{tbl:configuration},
minimal size of code clones 15 lines of code). As expected, if the distance value
increases, the recall increases with a simultaneous decrease in precision.
The precision values range between 96\% and 100\%. It is striking that with low
distance values, the recall values for Type 3 and Type 4 code clones are also lower.
In these cases, the distance value is lower than the admitted syntactical difference
of code clones. However, with increasing  distance values, more code clones of Type 3
and Type 4 can be found again, though with a higher inaccuracy. As these results
show, we can observe that recall and precision are balanced with a distance value
of $\tau=0.3$, i.e., a similarity threshold of $70\%$.

On a more technical side, we conduct further experiments for providing parameter
settings of StoneDetector for speeding up clone detection. First, in some cases,
two code fragments can be ruled out as code clones by simply analyzing the structure
of their respective description sets. For example, the factor in which the description
sets of two code fragments differ may be too large, or the lengths of two respective paths
in the code fragments' description sets may deviate too much. In such cases, the
comparison of the code fragments is short-circuited in StoneDetector for
speeding up clone detection. In order to determine adequate parameter values, we 
run multiple experiments on BigCloneBench using differing values for the
admitted maximal size factor of description sets and the admitted maximal
length difference of paths in description sets. 
Based on these experiments, we can derive that a factor of 1.7 as the maximal size
factor of description sets
and a maximum length difference of up to 7 nodes in paths to be compared yields the
best trade-off.

The abstraction of paths in StoneDetector, e.g., the uniform treatment
of variables and constants as discussed in Sect.~\ref{sec:approach}, can lead to several
identical paths in one description set. As an example, consider two paths that only differ
from each other in their last node, and where the respective nodes each represent 
assignments with which the same/different variables are assigned to different/the same
constants. By abstracting variables and constants, these assignments are then mapped to
the same representation, which results in an identical abstraction for both paths. We
observed such presence of identical paths in one and the same description set to occur
quite frequently. Although such occurrences do not influence StoneDetector's recall
or precision in finding code clones, its runtime is deteriorated due to the redundant
comparisons of identical paths. Therefore, in additional experiments, we compare a variant
of StoneDetector that does not handle identical paths and a variant where
two identical paths or two paths that differ only in one node are merged into a single
path. Surprisingly, this rather simple strategy already leads to a relative runtime
improvement of 20\%, which corresponds to a reduction from approx. 25 to 20 minutes on
the whole BigCloneBench.


\begin{table}
\centering
\footnotesize
\scalebox{0.8}{
    \begin{tabular}{l|c | c| c|c | c |c||c}
 & LCS with 4-& LCS$_{\mathrm{modified}}$ &Leven-
 &Needleman-&\multirow{2}{*}{Hamming}&Hamming & LCS$_{\mathrm{modified}}$ \\
 & byte hashing & with MD5 & shtein & Wunsch & & w/o penalty & with LSH \\ \hline
       Recall  &  20,831  &  20,828  & 20,828 & 20,828 & 20,828 & 20,828 & 20,828 \\
 T 1  & 99.9\%  & 99.9\%   & 99.9\%  & 99.9\%  & 99.9\%  &  99.9\% & 99.9\%  \\
  \hline
        Recall & 3,476 & 3,475 & 3,475 & 3,475 & 3,475 & 3,475  & 3,475 \\
   T 2  & 100.0\% & 99.9\% & 99.9\% & 99.9\% & 99.9\% & 99.9\%  & 99.9\% \\
  \hline
        Recall	& 3,272  & 3,270  & 3,270 & 3,270 & 2,849 & 2,889 & 3,209 \\
      VST 3  & 99.4\% & 99.4\% & 99.4\% & 99.4\% & 86.6\% & 87.8\% & 97.6\% \\
  \hline
        Recall & 6,702 & 6,773  & 6,404 & 6,721 & 4,406 & 4,666 & 6,808  \\
      ST 3  & 90.5\% & 91.5\% & 86.4\% & 90.7\%  & 59.5\% & 63.0\%  & 92.0\%    \\
  \hline
        Recall  & 4,335  & 5,727 & 3,312 & 4,956 & 1,255 & 2,253 & 6,324    \\
      MT 3  & 16.1\% & 21.4\%  & 12.3\% & 18.5\% & 4.4\% & 8.4\% & 23.6\%   \\
    \hline
       Recall  & 2,691  & 4,793  & 2,212 & 3,765 & 1,006 & 6,623 & 6,061 \\
   T 4   & 0.0\% & 0.1\% & 0.0\%  & 0.0\%   & 0.0\%  & 0.1\% & 0.1\%    \\
    \hline \hline
  \#Clones & 1,006,823 & 1,218,595 & 954,395 & 1,045,240 & 804,366 & 1,082,253 & 1,300,458 \\ \hline
  Precision & 99\% & 99\% & 99\% & 99\% & 99\% & 97\%  & 99\% 
    \end{tabular}
}
\captionof{table}{Recall of StoneDetector on Java source code on BigCloneBench using different string metrics and hashing methods
(T 1 - Type 1, T 2 - Type 2, VST 3 - Very-Strongly Type 3, ST 3 - Strongly Type 3, MT 3 - Moderately Type 3, T 4 - Type 4)}
\label{tbl:metrics}
\end{table}

Another parameter that can impact StoneDetector's capability in finding code clones
is the similarity measure, i.e., \emph{string metric}, used for comparing the paths of two
code fragments' dominator trees (cf. also Sect.~\ref{sec:metrics}). Furthermore,
since the path comparison is working on hashed fingerprints of the paths, the
utilized \emph{hashing mechanism} may as well influence StoneDetector's performance.
We therefore also want to specifically analyze the role of
the string metric and hashing mechanism. In our previously described experiments,
we used a combination of a variant of the longest common subsequence-based
metric ($\delta_{LCS_{modified}}$) in combination with Locality-Sensitive
Hashing (LSH). This is in contrast to the authors' previous work
in~\cite{DBLP:conf/icsm/AmmeHS21}, where we used the standard
LCS-based metric in combination with a simple 4-byte prime number-based
hashing mechanism. 

In a \textcolor{black}{third} experiment, we considered the role of using different string metrics
on clone detection in StoneDetector. To this end, we have chosen a subset of the
string metrics implemented in StoneDetector, i.e., the LCS-based metric,
modified LCS-based metric, Levenshtein distance, Needleman-Wunsch,
Hamming distance, and modified Hamming distance. The latter metric
depicts a variant of the Hamming distance in which different path lengths are
not penalized, in the sense that the difference in length between different paths is not included
in the distance measure as a penalty. We refer the interested reader to
Sect.~\ref{sec:metrics} for a short explanation of these string metrics. We analyze
the role of a certain string metric by evaluating StoneDetector's recall and
precision on BigCloneBench when using the respective metric and MD5 hashing,
keeping the other parameters stable (cf. Table~\ref{tbl:configuration}). Precision
is again measured by manual validation of a random sample.

The results of this \textcolor{black}{third} experiment are shown in Table~\ref{tbl:metrics}. Note
that we did not observe any differences in the results when using either MD5
or a simple 4-byte hashing and are therefore denoting the results for the standard
LCS-based metric with 4-byte hashing for the sake of comparison with our previous
work~\cite{DBLP:conf/icsm/AmmeHS21}. In summary, the shown results indicate that using
the Needleman-Wunsch algorithm or the modified LCS-based metric
achieves the best performance for clone detection. Compared with these metrics, the use
of the Levenshtein distance yields a visible deterioration in the recall values for
Moderately and Strongly Type 3 code clones, while precision remains the same. In contrast,
the Hamming distance implies significantly poorer recall values when looking at
the syntactically more deviating code clones, i.e., clones of Type 3 and Type 4.
While the use of the modified Hamming distance results in improved recall values,
it features the worst precision. \textcolor{black}{However, we observed significantly
shorter runtimes of StoneDetector in the case of using the Hamming distance,
as reported in Sect.~\ref{sec:scalability} and shown in Table~\ref{tbl:runtime}.}
As a rule of thumb to be taken from these measurements, we can thus state that the
Hamming distance should be preferred over the other metrics when searching exact
and near-miss clones, i.e., Type~1 and Type~2 code clones, due to its impact on
shortening StoneDetector's runtime.

In a \textcolor{black}{fourth} experiment, the hashing algorithm's influence on StoneDetector's
performance is evaluated. Since the modified LCS-based metric has so far achieved
the best results, we consider only this string metric in this experiment and
compare the tool's recall and precision when using either the MD5 hashing
algorithm or Locality-Sensitive Hashing (LSH). Note again that we did not
observe differences when using MD5 or a simple 4-byte hashing. As can also
be seen in Table~\ref{tbl:metrics}, the usage of LSH results in particular
in improved recall values for Moderately Type 3 and Type 4 code clones, while
maintaining precision. We have therefore chosen the modified
LCS-based metric and LSH as StoneDetector's default configuration in our
evaluation.
\bigskip \\
\noindent\fbox{
    \parbox{.95\textwidth}{
        \textcolor{black}{Clone detection on source code with StoneDetector achieves the best results using
        a similarity threshold of 70\% ($\tau=0.3$). Increasing the minimal
        size of clones in general increases the precision of clone detectors.   
        StoneDetector performs best using
        the modified longest common subsequence (LCS)-based string metric
        in combination with locality-sensitive hashing (LSH). Using the Hamming
        distance instead, though, allows for faster clone detection.}  
    }
} \bigskip

\subsection{\textcolor{black}{Research Question 5: Java Bytecode}}
\label{sec:bytecode}

\begin{table}
  \centering
 \footnotesize
\scalebox{0.85}{
    \begin{tabular}{l|c | c|c } 
 &\emph{StoneDetector (stack-based)}&\emph{StoneDetector (register-based)}&\emph{iClones (stack-based)} 
\\ \hline
       Recall    &  \textbf{20,792} & \textbf{20,792} & 20,777 \\ 
 T 1  & \textbf{99.9\%}  & \textbf{99.9\%}  & 99.9\%  \\ 
  \hline
        Recall  & 3,463 & \textbf{3,466} & 3,373 \\ 
   T 2 & 99.7\% & \textbf{99.8\%} & 97.2\% \\  
  \hline
        Recall	 & 3,157  & \textbf{3,238} & 3,043 \\ 
      VST 3 & 97.2\% & \textbf{99.7}\% & 93.7\% \\ 
  \hline
        Recall  & 5,035 & \textbf{6,430} & 3,559 \\ 
      ST 3 & 68.9\% & \textbf{88.0\%} & 48.7\% \\ 
  \hline
        Recall  & 2,804 & \textbf{4,311} & 1,345 \\ 
      MT 3 & 10.9\% & \textbf{16.8\%} & 5.2\% \\ 
    \hline
       Recall  & 3,178 & \textbf{3,754} & 1,087 \\ 
   T 4  & 0.0\%  & \textbf{0.0\%}  & 0.0\%  \\ 
   \hline \hline
  \#Clones & 439,767  & 600,522 & 362,403 \\ 
  Runtime & 77.20min  & \textbf{35.36min} & 37.59min \\ 
  Precision & 99\%  & 99\%  & \textbf{100\%} \\ 
  \end{tabular}
}
  \captionof{table}{Recall of clone detectors on stack-/register-based Java Bytecode representations on BigCloneBench 
(T 1 - Type 1, T 2 - Type 2, VST 3 - Very-Strongly Type 3, ST 3 - Strongly Type 3, MT 3 - Moderately Type 3,
 T 4 - Type 4)} 
  \label{tbl:bytecode}
\end{table}

For our last research question, we consider Java's Bytecode representation for
clone detection. In general, we are interested in whether conventional code clone detectors
like StoneDetector can be used for finding source code clones in their Java Bytecode
representation. This use case may become apparent in situations where the source code
is not available, e.g., when analyzing compiled Java code with regard to included malware.

Clone detection for Java Bytecode can be done either directly using the raw Bytecode or
using an intermediate representation as generated, e.g., during compilation.
The Soot framework~\cite{DBLP:conf/cascon/Vallee-RaiCGHLS99} supports
the transformation of Java Bytecode into several intermediate representations.
Soot's \emph{Baf} format roughly resembles the original Bytecode and provides a stack-based
intermediate representation. In contrast, the \emph{Jimple} format is a register-based
three-address code, which is generated by Soot out of the stack-based Bytecode for
program analysis and transformation. We consider both formats in our evaluation and
are interested in their different suitability for code clone detection.  
Evaluating clone detectors for Java Bytecode is unfortunately not as straightforward
as for Java source code, because the state-of-the-art benchmark BigCloneBench
consists of only plain source files and is missing their respective
dependencies~\cite{DBLP:conf/iwsc/SchaferHA23}. We therefore use the
\emph{Stubber} tooling to transform BigCloneBench's Java source files in such a way,
that they can be compiled to Java Bytecode without the missing dependencies. For the
details, we refer the interested reader to the publication introducing
Stubber~\cite{DBLP:conf/iwsc/SchaferAH21}. Using Stubber allows us to generate
Bytecode for 52,816 or 95\% of the overall 54,999 benchmark's source files
and to subsequently generate the Bytecode's Baf and Jimple representation using Soot.

Due to characteristics of Java Bytecode as well as the Baf and Jimple formats,
StoneDetector applies certain heuristics for eliminating some kind of low-level
instructions that are specific to the instruction format. For example,
with respect to the Baf format, there are several load and store instructions for
pushing and pulling values onto and from the stack. As these instructions have
no matching counterparts in the source code, they are eliminated before clone
detection is performed in StoneDetector. Similar heuristics are applied to
eliminate the duplicate paths in a dominator tree, which have been introduced in
Jimple and Baf for the representation of the source code's high-level control
flow structures~\cite{DBLP:conf/iwsc/SchaferHA23}. Furthermore, the LCS-based metric
with 4-byte hashing is used, while code clones are again
limited to code fragments with equal to or more than 15 lines of code. For the
Baf format, $\tau = 0.15$ is chosen, and for the Jimple format, again $\tau = 0.3$ is
used.

We also compare StoneDetector's performance on Java Bytecode with another
state-of-the-art clone detector, i.e., iClones. The tool does not readily come with
support for Bytecode and we have thus extended iClones for the stack-based Java
Bytecode. The results of our experiments on Java Bytecode are shown in
Table~\ref{tbl:bytecode}. As can be seen, StoneDetector can be effectively used as
a clone detector on Java Bytecode and provides a better performance with respect
to recall and precision when compared to iClones.
Note, though, that the
latter 
tool has not been designed for Java Bytecode or other compiled
code artifacts. We also note that StoneDetector performs better on the
register-based Jimple format compared to the stack-based Baf format.
This is justified
by the fact that in the Bytecode, smaller changes in the original Java source code lead to larger
differences in the intermediate format due to the characteristics of the instruction set
and underlying stack machine model. While this also influences StoneDetector's performance
with Jimple, the deteriorating effects are larger for the Baf format.
This plays no role with respect to exact or
near-miss code clones of Type 1 and 2, as can also be seen in Table~\ref{tbl:bytecode},
but becomes more of an issue for Type 3 and 4 clones with a larger proportion
of deviating source code instructions.
\bigskip \\
\noindent\fbox{
    \parbox{.95\textwidth}{
        Detecting source code clones based on Java Bytecode is in general possible.
        A register-based representation of Java Bytecode is better suited
        for their detection compared to a stack-based representation.
    }
} \bigskip

\subsection{\textcolor{black}{Summarization}}

\textcolor{black}{Our evaluation of StoneDetector in comparison with state-of-the-art conventional
clone detection tools illustrates its strengths and its limitations. While
near-miss clone detectors like NiCad, SourcererCC, and CloneWorks excel in
identifying Type 1 and Type 2 code clones with high precision and fast runtimes
(cf. Sect.~\ref{sec:bigclonebench}, Sect.~\ref{sec:scalability}), StoneDetector
achieves in its default configuration competitive performance results, though not
providing for similar fast clone detection runtimes. StoneDetector however demonstrates its capabilites
in finding harder-to-detect clones of Type 3 and Type 4 with acceptable precision in
reasonable time, also when compared to advanced and/or Deep Learning tools such as
NIL and Oreo, as exemplified on benchmarks Google Code Jam and Project CodeNet where
StoneDetector achieved the best overall recall (cf. Sect.~\ref{sec:otherbenchmarks}).
The manual qualitative examination of code clones revealed that the tool in fact finds
certain types which are not detected by any of the other tools (cf. Sect~\ref{sec:examination}). Furthermore,
StoneDetector can be adapted to other configurations (cf. Sect.~\ref{sec:settings}),
e.g., using the Hamming distance metric for scalable and fast near-miss clone detection,
which makes it a versatile and rich platform for Java code clone detection and experimentation,
including Bytecode-based clone detection (cf. Sect.~\ref{sec:bytecode}).} 

\textcolor{black}{On the one hand, StoneDetector in its default configuration is thus
most beneficial in software maintenance scenarios, where developers are challenged
by growing inconsistencies across clone instances~\cite{DBLP:conf/msr/XieKZ13}.
Note that such syntactically more deviating clone instances have also been shown to be
prone to delays in propagating code changes, referred to as late
propagation~\cite{DBLP:journals/sqj/MondalRS16}, potentially resulting in bugs,
which makes their detection and tracking even more important. In these cases,
StoneDetector's tolerance for refactored statements, i.e., structural code clones, and
larger code edits or subsumed/embedded code fragments, i.e., subclones, enable it to capture
clones and therefore bugs that other clone detectors may miss, as revealed
in our evaluation. 
Such scenarios are especially relevant in software maintenance, where inconsistent
and insufficient identification of code clones has been shown to correlate with
fault proneness~\cite{DBLP:journals/ijseke/IslamMRS17}.
StoneDetector's ability to detect structurally modified code clones also supports
consolidation of functional and architectural
adaptations~\cite{DBLP:journals/tosem/AssiHZ25}, which is less feasible with clone
detectors optimized for exact or near-miss clones. On the other hand, StoneDetector
allows for other configurations. For example, in certain scenarios, the primary
objective of clone detection may be the rapid identification of exact duplicates
or near-miss clones in large codebases. Note that such code clones often
constitute a large fraction in practice~\cite{DBLP:journals/pacmpl/LopesMMSYZSV17}.
As mentioned above, using the Hamming distance metric and MD5 hashing makes
StoneDetector behave like a fast near-miss clone detector, yielding a speed-up
of approx. 6.3, though trading 
accuracy for speed. Apart from StoneDetectur thus being adaptable to
software maintenance scenarios, the tool provides a rich platform for Java
code clone experimentation including various configuration
parameters like, e.g., used distance metrics,
hashing algorithms, code representations.}

\section{Related Work} \label{sec:relatedwork}

Techniques and tools for clone detection address the problem of finding
code fragments that share a certain degree of syntactic and/or semantic
similarity. Approaches for the detection of code clones can be classified
according to the used detection
methodology~\cite{DBLP:journals/jss/NasrabadiPRRE23}.
Following a recent survey on this research topic
by~\citet{DBLP:journals/jss/NasrabadiPRRE23}, we differentiate text-, token-,
tree-, graph-, and metric-based methods as well as hybrid, \textcolor{black}{Machine/Deep Learning}, and
dynamic approaches. For an extensive discussion of code clone techniques and tools,
we refer the interested reader to one of the various surveys available on the
topic~\cite{DBLP:journals/scp/RoyCK09,DBLP:journals/jss/NasrabadiPRRE23,DBLP:journals/infsof/RattanBS13,DBLP:journals/access/AinBAAM19}
and here provide a condensed overview of the most formative and relevant work.

\subsection{Text-based Techniques}

The raw source code is used as input for the first group of methods. The
source code is either used directly as a sequence of strings for finding
matching parts of two code fragments, or only very limited pre-processing
is applied beforehand, e.g., removal of whitespace and/or code normalization.
Despite their simplicity and limited effectiveness on Type 3/4 code clones,
text-based methods are still used today due to their efficiency and easy
use for different programming languages. Examples include
the general-purpose clone detectors like
\textit{Simian}\footnote{\url{https://simian.quandarypeak.com/}} or
plagiarism checkers like
\textit{MOSS}~\cite{DBLP:conf/sigmod/SchleimerWA03}\footnote{\url{https://theory.stanford.edu/~aiken/moss/}}.
As one of the first approaches, Baker proposed the tool
\textit{Dup}~\cite{DBLP:conf/wcre/Baker95}.
The tool combines a consistent identifier/literal renaming scheme with
suffix trees on hashed code lines for finding parametrized
code matches, i.e., near-miss code clones. Similarly, Johnson proposed
to use the Karp-Rabin algorithm and sliding windows for hashed code lines,
called fingerprints, to find duplicated code
fragments~\cite{DBLP:conf/cascon/Johnson93}.
In order to also detect Type 2 clones, he also removes whitespace and
comments and maps all identifiers, keywords, and literals to a
common placeholder.
The text-based clone detector \textit{Vuddy} uses whole-function
hashes on pre-processed source code for generating function
fingerprints. These fingerprints are then used as indexes for the
implementation of efficient lookups for vulnerable code in large
code bases~\cite{DBLP:conf/sp/KimWLO17}. 
A well-known and up-to-date clone detector is \emph{NiCad}~\cite{DBLP:conf/iwpc/RoyC08a,DBLP:books/sp/21/MondalRC21},
which is mostly text-based
but also applies lightweight parsing for normalizing the raw
source code so that noise is removed and code formats are
standardized. The code pre-processing can be adjusted to
different programming languages using a customizable grammar
definition.
NiCad is using the longest common subsequence algorithm
for finding matching code and is also one of the
clone detectors used in our comparative evaluation in
Sect.~\ref{sec:evaluation}. The tool identifies almost
all Type 1/2 clones and a substantial part of Type 3 code
clones~\cite{DBLP:conf/icsm/SvajlenkoR15}.
Various other authors have used NiCad as a starting point for
their own research, e.g.,~\citet{DBLP:conf/iwsc/Ragkhitwetsagul17} applied
NiCad to compiled and subsequently decompiled code for further
code normalization, while~\citet{DBLP:journals/jcst/ChenADZ15} used NiCad on decompiled
Bytecode to identify Android malware.

\subsection{Token-based Techniques}

Token-based clone detectors use a lexer and/or parser for transforming
raw source code into token sequences, which are then further analyzed
to find matching code fragments. In this way, minor code edits, e.g.,
renaming of variables, can be easily handled. While token-based tools
have in general shown good performance in the detection of near-miss
code clones, more recent tools also provide good capabilities in the
detection of large-gap Type 3 code clones featuring many code edits~\cite{DBLP:conf/icse/WangSWXR18,DBLP:journals/access/LiuWFWL19,DBLP:journals/access/WuWYCXR20}. Furthermore, modern token-based clone detectors typically
scale up to large code bases with several hundred million
lines of code~\cite{DBLP:conf/icse/SajnaniSSRL16}. 

\emph{CCFinder}~\cite{DBLP:journals/tse/KamiyaKI02} lexes two source
code fragments into token sequences, thereby transforming some tokens
according to some rules, e.g., replacing mapping identifiers to a
common token. The thus normalized token sequences are then handed
over to the suffix tree algorithm for finding matching code fragments.
Note that, as some of the above-mentioned text-based tools work quite
similarly, the differentiation between text- and token-based approaches
to clone detection is rather not strict. In subsequent work, the
authors provided a flexible tokenization mechanism for easily
extending the tool towards additional programming
languages~\cite{DBLP:conf/apsec/SemuraYCI17}. The suffix tree in
its generalized form is also used by G\"ode and Koschke in their
tool \emph{iClones} for incremental code clone detection,
i.e., detection of code clones in evolving code
bases~\cite{DBLP:conf/csmr/GodeK09}. \emph{CP-Miner} uses
frequent subsequence mining for finding interleaved token
subsequences in case of reordered statements or inserted code
within code lines~\cite{DBLP:journals/tse/LiLMZ06}. 
The token-based clone detector \emph{SourcererCC} uses
optimized indexing and filtering techniques for reducing the
number of candidate code fragments when searching for code clones. 
In this way, the tool supports scalable clone detection for
large code repositories with hundreds of millions of lines of
code~\cite{DBLP:conf/icse/SajnaniSSRL16}.
Scaling clone detectors for large codebases has also been 
the focus of subsequent research~\cite{DBLP:conf/wcre/LiWRSPZHM20}.
\textcolor{black}{\emph{CloneWorks} also uses a token-based approach combined
with several techniques to scale and fasten clone detection, 
including Jaccard similarity metric, sub-block filtering heuristics,
index creation and partitioning~\cite{DBLP:conf/icse/SvajlenkoR17a}.}
Another token-based technique is described
by~\citet{DBLP:journals/ese/Ragkhitwetsagul19}, their tool
\emph{Siamese} uses token sequences with different levels
of abstractions and methods from document retrieval on their
n-gram vector representations. As a result, they are able to
provide a scalable method for the detection of Type 1, 2, and 3
code clones.  

While tools like NiCad and SourcererCC are quite good at finding exact
and near-miss clones, Type 3 and 4 clones with lower syntactical similarity
provide a challenge. More recent token-based clone detectors,
like \emph{CCAligner}~\cite{DBLP:conf/icse/WangSWXR18},
\emph{LVMapper}~\cite{DBLP:journals/access/WuWYCXR20}, and
\emph{NIL}~\cite{DBLP:conf/sigsoft/NakagawaHK21},
therefore focus on code clones with many consecutive code edits
or many code modifications scattered around the
code. These so-called large-gap clone detectors often
 apply more sophisticated string metrics for aligning and matching token
sequences, e.g., e-edit distance~\cite{DBLP:conf/icse/WangSWXR18}
or Smith-Waterman sequence alignment~\cite{DBLP:journals/access/LiuWFWL19}.
We include the tools \emph{SourcererCC}, \emph{iClones}, \emph{CloneWorks},
\emph{NIL}, and \emph{CCAligner} as representative token-based
clone detectors in our evaluation in Sect.~\ref{sec:evaluation}

\begin{table}
    \footnotesize
    \begin{tabular}{c|c|c|c|c|c}
      \rotatebox{90}{Approach}\rotatebox{90}{based on} & \rotatebox{90}{Supported}\rotatebox{90}{clone types} & \rotatebox{90}{Accuracy} &  \rotatebox{90}{Efficiency} & \rotatebox{90}{Strength} & \rotatebox{90}{Weakness} \\ \hline
      
    Text  & 1, 2, part of 3 & + & + & language-agnostic & fails on Type 3/4 \\ \hline
    Token  & 1, 2, part of 3 & + & + & good scalability & fails on Type 3/4 \\ \hline
    Tree/ & \multirow{2}{*}{1, 2, 3, 4} 
   & \multirow{2}{*}{+} & \multirow{2}{*}{-} & uses syntactical and &   construction and \\
    Graph&& &  & semantic information & matching expensive \\ \hline
    Metric  & 1, 2, 3, 4 & + & $\circ$ & language-agnostic & threshold sensitivity \\ \hline
    Deep  & \multirow{2}{*}{1, 2, 3, 4} & \multirow{2}{*}{+} & \multirow{2}{*}{-} & fully automated, & requires training, \\
    learning& & & & very flexible & prone to overfitting \\ \hline
    Dynamic & 1, 2, 3, 4 & $\circ$ & - & finds semantic clones & specialized for use case \\ \hline
    Hybrid  & \multicolumn{5}{c}{depends on technique}
    \end{tabular}
    
   \caption{Comparison of code clone detection techniques (- low, $\circ$ adequate, + high)}
    
    \label{relatedwork}
\end{table}

\subsection{Tree- and graph-based Techniques}

Tree- or graph-based techniques parse the source code
into abstract syntax trees or use even further transformations for generating
control flow or program dependence graphs. While abstract syntax trees
contain all of the syntactical information about code fragments, edges
in control flow graphs depict control dependences and in program dependence
graphs control as well as data dependences and thus convey the semantics
of code fragments. Thus, both approaches are able to detect code clones of
Type 3 and Type 4, which share the same semantics but use differing syntax.
The similarity matching for code fragments in tree- or graph-based techniques
is performed on the generated program representations by finding subtrees
or looking for graph isomorphisms. Both, the generation of the program
representation as well as the tree or graph matching is often computationally
expensive. 

As one of the early tree-based clone detectors, \emph{CloneDR} computes
hash values for abstract syntax trees, which allows lowering the number
of expensive tree comparisons, as only abstract syntax trees with the
same hash value need to be compared~\cite{DBLP:conf/icsm/BaxterYMSB98}.
More permissive tree comparisons are used by~\citet{DBLP:conf/wcre/EvansFM07},
where they apply, besides lexical abstractions, matching, e.g., differing
identifiers, also structural abstractions, where whole subexpressions are
matched, for finding more code clones on abstract syntax trees.  
\emph{Deckard} is another tree-based tool, which uses the tree edit
distance for matching code fragments based upon their representation
as abstract syntax trees. In comparative evaluations, Deckard has
been shown to find more Type 3/4 code clones, but at considerably
lower precision~\cite{DBLP:conf/icse/WangSWXR18,DBLP:conf/sigsoft/SainiFLBL18},
a result we can reproduce in our evaluation in Sect.~\ref{sec:evaluation}.
Sager et al. use a combination of common subtree isomorphism and tree edit
distance for detecting similar code fragments based upon a representation
derived from abstract syntax trees~\cite{DBLP:conf/msr/SagerBPK06}.
Yang et al. linearize abstract syntax trees and apply the Smith-Waterman
sequence alignment algorithm to the linearized trees for
finding similar code fragments and thus code clones~\cite{DBLP:conf/compsac/YangRCJ18}.

Komondoor and Horwitz present their clone detector \emph{PDG-DUP}~\cite{DBLP:conf/sas/KomondoorH01}, where they use backward slicing
on program dependence graphs for finding isomorphic subgraphs and thus
code clones. While the slicing approach helps in increasing efficiency,
the technique is still time-consuming for programs with large program
dependence graphs. Krinke introduces \emph{duplix} as another tool
working with program dependence graphs, which iteratively searches
for maximal similar subgraphs~\cite{DBLP:conf/wcre/Krinke01}, but
also notes the computational complexity of the approach. A similar 
approach based on isomorphic subgraph matching has been implemented
 for plagiarism detection~\cite{DBLP:conf/kdd/LiuCHY06}. Another
approach, where slices on program dependence graphs are hashed and
compared
via MD5 or Locality-Sensitive Hashing (LSH)
is implemented in the tool \emph{srcClone}~\cite{alomart2020}.
Breaking
code into smaller code fragments, which are mapped to program
dependence graphs, and using filtering based on different features
of program dependence graphs before subgraph isomorphism testing
allows \emph{CCSharp} to lower the computation
burden~\cite{DBLP:conf/apsec/WangWX17}. A similar multi-stage
approach is described by~\citet{DBLP:journals/pcs/SargsyanKBA16}
using the LLVM framework for program dependence graph generation.
Approximative graph matching using graph kernels is applied by
the clone detector \emph{CCGraph}~\cite{DBLP:conf/kbse/ZouBXX20}
for speeding clone detection up based upon program dependence graphs.
Other graph-based program representations besides program dependence
graphs are rather used sporadically
for clone detection, e.g., call graphs~\cite{DBLP:conf/ccs/HuCS09}.
Using dominator trees for clone detection has been presented by
the authors of this paper previously~\cite{DBLP:conf/icsm/AmmeHS21,DBLP:conf/acsos/SchaferAH20}.
The authors are not aware of any other research using dominator
trees or dominator information for code clone detection.

\subsection{Metric-based Techniques}

A straightforward approach for finding similar code clones can be
implemented by collecting characteristics of code fragments
and matching code fragments with similar characteristics
according to some variance threshold. Several metrics can be used
for characterizing code fragments, including, e.g., the number 
of lines of code, the number of used variables and arguments,
cyclomatic complexity, the number of various types of statements,
and so on. While metric-based clone detectors are rather straightforward
to implement and adjustable to different programming languages, they are
quite sensitive to choosing the right threshold for similarity measurement. 
Further note that metric-based approaches can be seen as very simple
data-driven approaches, since adding a mapping of the metrics into
a vector space and an optimizable algorithm for distinguishing
code clones from non-code clones based on this vector space
yields the latter techniques~\cite{DBLP:conf/kbse/NafiKRRS19}.
In 1996, Mayrand et al. described a first approach for detecting
code clones based upon a set of several metrics, including aspects
like identifiers,
layout, expressions, and control flow~\cite{DBLP:conf/icsm/MayrandLM96}.
Similar function-level metrics, such as number of statements,
McCabe’s cyclomatic complexity, and number of use-definition pairs,
are used for clone detection by~\citet{DBLP:conf/iwpc/PatenaudeMDL99}. 
Metrics like type and order of statements, type and number of operators,
and number of variables are also
used by~\citet{DBLP:journals/eswa/SudhamaniR19}
to build so-called feature
tables for code fragments, which can then be compared using distance
metrics like the city block metric. The technique showed reasonable
results for Type 1, 2, and part of Type 3 clones.
With the advent of multi-stage code clone detectors, metric-based
techniques have also been proposed for pre-filtering candidate
pairs of code fragments according to their metrics~\cite{DBLP:conf/sigsoft/SainiFLBL18,DBLP:conf/dsa/JinCLZ22}.

\subsection{\textcolor{black}{Machine/Deep Learning Techniques}}

In the last decade, the application of machine learning and Deep Learning
has become more widespread in clone detection. Common to these approaches
is the embedding of code fragments into a manually crafted or automatically
learned and often high-dimensional vector space, wherein statistical
relationships are used for predicting similar code. The ability to thus
automatically learn and recognize the hidden patterns of code clones from,
e.g., today's large and omnipresent code repositories, should guarantee
promising results. While in particular Deep Learning clone detectors have
shown very good performance in the detection of Type 1 to Type 4 code clones
lately, research has recently also highlighted common problems and pitfalls
in the application of Deep Learning to clone
detection, i.e., training/evaluation bias, overfitting, and missing ground
truth~\cite{DBLP:journals/corr/abs-2505-04311,DBLP:conf/iwsc/SchaferAH22,DBLP:conf/iwsc/KrinkeR22,DBLP:conf/iwpc/ChoiFFYI23}. 

Prior approaches considered conventional machine learning methods. For instance,
Cesare et al. use algorithms like naive Bayes and random forests for detecting
package-level clones~\cite{DBLP:conf/securecomm/CesareXZ13}.
Clustering via k-Means can also be used for
unsupervised learning of similar code and for finding code clones with respect
to various similarity thresholds~\cite{DBLP:conf/wcre/Keivanloo0Z15}.
More recent work, however, has shifted towards
the application of deep neural networks.
In pioneering research, White et al. use an autoencoder architecture for
unsupervised learning of abstract syntax trees and compute code similarities
on the learned representation~\cite{DBLP:conf/kbse/WhiteTVP16}. The Deep Learning
clone detector \emph{CCLearner} uses the frequencies of token categories 
in the abstract syntax trees of code fragments as features and learns to
discriminate code clones from non-code clones using these features via
supervised training of a deep neural network~\cite{DBLP:conf/icsm/LiFZMR17}. 
\emph{CDLH}~\cite{DBLP:conf/ijcai/WeiL17} is using the long short-term
memory (LSTM) architecture for representation learning on abstract syntax
trees and is trained to embed the resulting representation to hash values,
such that code clones are close in terms of their Hamming distance. Similarly,
Zhang et al. are using in their architecture \emph{ASTNN} bidirectional
recurrent units on embedded statement trees for encoding abstract syntax
trees and identifying code
clones~\cite{DBLP:conf/icse/ZhangWZ0WL19}.
Different settings for embeddings
of abstract syntax trees are evaluated by~\citet{DBLP:conf/wcre/BuchA19},
where they use a
siamese architecture that optimizes the embeddings of two abstract
syntax trees such that the cosine similarity in the case of code clones is
maximized. As an interesting result, pre-trained embeddings
like~\emph{Word2vec, Code2vec}~\cite{DBLP:journals/pacmpl/AlonZLY19} have proven
useful. Another lightweight approach for embedding abstract syntax
trees is presented by~\citet{DBLP:conf/icsm/GaoWLYSC19}, where the authors
use random walks on the abstract syntax trees and embed the resulting token sequences similar to \emph{Word2vec}.
The resulting encodings of code fragments
can then be compared via their Euclidean distances.
\textcolor{black}{Hu et al.~\cite{DBLP:conf/kbse/HuZPWS022} choose manual
feature engineering to encode abstract syntax trees using different centrality
values and then employ decision trees for finding matching code clones
based on the resulting vector representations. Different
similarity measures based upon code fragments' token sequences are
considered by Feng et al.~\cite{DBLP:conf/icse/FengSWZ0024},
which feed the resulting feature vectors into several classical
machine learning algorithms for detecting code clones.}

The Siamese architecture is also used in the hybrid clone detector
\emph{Oreo}, where the model is used with metrics-based
feature vectors for detecting harder-to-find clones,
achieving in particular good results for Moderately Type 3 clones and
beyond~\cite{DBLP:conf/sigsoft/SainiFLBL18}. We use Oreo
as a representative for \textcolor{black}{Deep Learning} clone detectors in
our evaluation in Sect.~\ref{sec:evaluation}.

The more specialized Deep Learning architecture \emph{FA-AST-GMN} for the
detection of code clones based on graph matching networks~\cite{DBLP:conf/icml/LiGDVK19}
has been introduced for abstract syntax trees~\cite{DBLP:conf/wcre/WangLM0J20}.
Furthermore, other code representations besides abstract syntax
trees have also been considered as input for Deep Learning clone
detectors~\cite{DBLP:conf/ccs/XuLFYSS17,DBLP:conf/sigsoft/ZhaoH18}.
Control and data flow is combined into a rich feature model
in~\emph{DeepSim} for training a Deep Learning model for clone
detection~\cite{DBLP:conf/sigsoft/ZhaoH18}. Fang et al. combine
abstract syntax trees and control flow graphs in so-called fusion
embeddings for finding in particular Type 4 code
clones~\cite{DBLP:conf/issta/FangLS0S20}.
Patel and Sinah~\cite{DBLP:conf/pts/PatelS21} combine this approach with the
Siamese architecture. The usage of the different code
representations: identifiers, abstract syntax trees, control
flow graphs, and Bytecode, as well as their combined usage for
clone detection is further
studied by~\citet{DBLP:conf/msr/TufanoWBPWP08}. Various code
metrics and embeddings of API calls have also been used as input features
to a siamese network for cross-language clone detection~\cite{DBLP:conf/kbse/NafiKRRS19}.
Finally, pre-trained models based on the transformer
architecture, e.g., \emph{CodeBERT}~\cite{DBLP:conf/emnlp/FengGTDFGS0LJZ20},
\emph{GraphCodeBERT}~\cite{DBLP:conf/iclr/GuoRLFT0ZDSFTDC21}, and even large language 
models like \emph{GPT}~\cite{DBLP:journals/corr/abs-2308-01191} have been
evaluated and adjusted for clone
detection~\cite{DBLP:conf/iwpc/TaoZHX22,DBLP:conf/icsm/ChochlovAPLHGB22}.
Note that this paper focuses on conventional approaches to code clone
detection, which the authors believe to play a role in the future.
 We, though, acknowledge the potential of such methods and
are interested in the in-depth analysis of their strengths
and weaknesses in future work.

\subsection{Dynamic Techniques}

Dynamic methods focus on the computations of code instead of its
syntax and usually target binary code. Therefore,
dynamic analyses or testing is used to execute code fragments
with sample inputs and track their behavior.
The resulting 
data can then be used to match semantically similar code and,
in this way, in particular, to detect Type 4 code clones.
As for all dynamic methods,
scalability and execution time can be problematic,
and accuracy is tightly coupled to the coverage of test inputs.

Pewny et al. identify semantic code clones in binary code by 
first translating the binary code into a common intermediate
representation, which is then interpreted block-wise for
random input values. The resulting input/output pairs of
code fragments are hashed and the semantic hashes can
afterward be used for finding semantically similar
code~\cite{DBLP:conf/sp/PewnyGGRH15}. Similar approaches are
chosen by~\citet{DBLP:conf/iwpc/HuZLG17}, where behavior
of code fragments is emulated, and by~\citet{DBLP:conf/issta/JiangS09},
where code fragments are instrumented and executed on sample
inputs for tracking input/output behavior.
While the above-mentioned approaches consider compiled
C/C++ binaries, other techniques have been proposed for
Java as well~\cite{DBLP:conf/iwpc/SuBKS16,DBLP:conf/iwsc/LeoneT22}. 
Instrumentation of binary code is also used by~\citet{DBLP:conf/sigsoft/SuBHSKJ16} for generating
dynamic dependency graphs, where semantically code
fragments are discovered via searching isomorphic
subgraphs. Alternatively, matching semantic
clones can be done done using the notion of abstract memory
states, which are derived by
a static analyzer and compared based on the abstract
states' semantics~\cite{DBLP:conf/icse/KimJKY11}.
In this way, clone detection is again aligned
with the semantics of code fragments and
independent of syntactical similarity. 
Due to the different focus, i.e., heterologously instead of
homologously developed code, and specialized use case of these
dynamic approaches, we do not include them in our evaluation.

\subsection{Hybrid Techniques}

Hybrid techniques combine the aforementioned approaches 
for aggregating their advantages. For example, as can also be seen in
Table~\ref{relatedwork}, textual or token-based techniques usually
suffer from lower effectivity in identifying Type 3 and Type 4 code clones
with larger syntactical variance, and tree- or graph-based techniques
demonstrate increased effectivity in finding harder-to-detect code clones
but are mostly less efficient. Thus, a combination of both should provide
an effective and efficient code clone detector. Besides, we frequently
also found combinations of text- and token-based techniques as well as
token- or tree-based techniques with Deep Learning approaches,
i.e., data-driven techniques.
Since distinguishing the conventional approaches and hybrid approaches
cannot, as a matter of fact, be strict, some hybrid techniques have already
been discussed  before. For example,
Nicad~\cite{DBLP:conf/iwpc/RoyC08a,DBLP:books/sp/21/MondalRC21}
combines a basically textual approach
to clone detection with lightweight and token-based preprocessing,
Yang et al. generate the abstract syntax tree of code fragments for
encoding structural information but then compare the linearized trees
via the token-based sequence alignment~\cite{DBLP:conf/compsac/YangRCJ18},
and Oreo~\cite{DBLP:conf/sigsoft/SainiFLBL18} uses a two-step approach by using
a metric-based technique as a basis for finding code clones and a Deep
Learning model for the remaining harder-to-detect code clones.  
In the following, we present some other representative examples which fit
into the category of hybrid approaches.

Token-based and tree- or graph-based approaches to clone detection are often
used in such a way that a token-based technique is used for filtering code clone
candidates which are afterward validated using the tree-/graph-based
technique~\cite{DBLP:journals/scp/ChilowiczDR13}.
A combination of token- and tree-based clone detection
is also used by~\citet{DBLP:conf/wcre/VislavskiRCB18} in their
tool \emph{LICCA}
 to detect cross-language
code clones and by~\citet{DBLP:journals/jss/WangDXX23}
in their tool \emph{CCStokener} to speed up detection efficiency while
retaining detection performance. Similar two-step pipelines have been presented
for metric-based approaches, where the metric-based technique is used for
prefiltering clone candidates and a graph- or text-based technique is used
for validation~\cite{DBLP:journals/fcsc/WangWSM14,DBLP:journals/jserd/KodhaiK14}.
Further research focused on the role of different intermediate representations
for code clone detection, e.g., considering Java source code and Bytecode.
Selim et al. apply the clone detector CCFinder~\cite{DBLP:journals/tse/KamiyaKI02}
on the Jimple representation
of Java Bytecode and find source code clones that are not 
detected otherwise. They thus propose to use their approach to complement
clone detectors working solely on the source code~\cite{DBLP:conf/wcre/SelimFZ10}.
A similar method has been described for the usage of NiCad~\cite{DBLP:conf/iwpc/RoyC08a}
on LLVM-based intermediate representations for C~\cite{DBLP:conf/iwsc/CaldeiraSWFS20}.
A more comprehensive approach, including a study of token-, tree-, and graph-
based code representations and a combination thereof within the tool \emph{TACC} is
presented by~\citet{DBLP:conf/icse/WangYWZXL23},
where they demonstrate an increased performance
for the combined approach for all types of code clones. The usage of different
code representations in combination with Deep Learning is
also analyzed by~\citet{DBLP:conf/msr/TufanoWBPWP08}.

\section{Conclusion} \label{sec:conclusion}

In this paper, we have presented an approach to code clone detection
for the Java programming language using dominator trees and approximate
string matching. Using dominator trees allows us to take advantage of
control flow information in order to find harder-to-detect code clones
with many syntactical deviations or semantic clones, which feature
low syntactical similarity but share the same computations. The clone
matching itself is, though, based upon string metrics, which are applied
to encoded paths of the dominator trees for exploiting the advantages
of conventional clone detecting, in particular in the case of exact and
near-miss code clones. In this way, our method is different from most
other conventional approaches to clone detection for Java, since the
approach incorporates control flow besides syntax and, as a result,
achieves better performance when asked to find code clones with
larger syntactical variance, i.e., Type 3 and Type 4 clones.

The method for clone detection has been
implemented in the publicly available \textcolor{black}{StoneDetector tool}, which
provides a versatile system for finding code clones in both Java
source code and Bytecode, as well as supporting manifold experiments
on clone detection's parameters. Using the implementation
in StoneDetector, we conducted a thorough evaluation of our approach
with respect to the state-of-the-art benchmarks for Java code clone detection,
i.e., BigCloneBench, Google Code Jam, GPTCloneBench, and Project CodeNet.
Furthermore, we experimentally compared the performance of the code clone
detection approach in StoneDetector to other state-of-the-art clone
detectors, including NiCad, iClones, CCAligner, SourcererCC, Deckard,
Oreo, \textcolor{black}{CloneWorks} and NIL. The experiments proved StoneDetector's competitive
capabilities in finding exact near-miss clones with respect to precision
and recall. In addition, StoneDetector excels at identifying code clones
with larger syntactical variance at high precision when compared to
the other state-of-the-art tools. Further experiments demonstrated
StoneDetector's ability to analyze large code bases with up to
multiple million lines of code. Besides its performance, we highlight
the versatility of \textcolor{black}{StoneDetector} experimenting with different
parameters for code clone detection, including the
similarity threshold, minimal size of code clones,
string metrics, e.g., Hamming distance, Longest Common Subsequence,
and hashing methods, e.g.,
Locality-Sensitive Hashing. Eventually, StoneDetector's ready-made extensibility to other programming
languages and program representations is presented, as is shown for
stack-based and register-based Bytecode formats.

While this paper focuses on conventional code clone detection and
therefore omits an in-depth discussion and comparative evaluation
with Deep Learning approaches to code clone detection, we want to
analyze and incorporate such models into \textcolor{black}{StoneDetector}
in our future work. While we acknowledge the promising results of,
in particular, large language models in recent work~\cite{DBLP:journals/corr/abs-2308-01191},
as discussed in the related work section, we also assume that conventional
methods of clone detection conjoined with their advantages, as
presented in this paper, will play a role in the future. We though
concede that the assessment of the advantages and disadvantages
of conventional code clone detection in comparison with Deep Learning
approaches, as well as their incorporation into a possible combined
approach, requires further research and analysis. Besides, we are 
also interested in extending \textcolor{black}{StoneDetector} with
respect to other programming languages, which have not been in
the research focus for code clone detection. \textcolor{black}{Finally,
we believe that StoneDetector's runtime can be improved by
further optimizing its implementation as well as examining
configurations which trade off accuracy against better runtime, as
has been discussed for the Hamming distance.} 
 


\bibliographystyle{elsarticle-num-names}
\bibliography{references}

\end{document}